\documentstyle[12pt,epsfig,subfigure,rotating,floatflt,glas,ssi]{article}
\newlength{\dinwidth}
\newlength{\dinmargin}
\input epsf
\epsfverbosetrue

\def\st{\scriptstyle}

\def\be{\begin{equation}}
\def\ee{\end{equation}}
\def\bea{\begin{eqnarray}}
\def\eea{\end{eqnarray}}

\newcommand{\do}{D^0}

\newcommand{\pom}{I\!\!P}

\newcommand{\xpom}{x_{\pom}}

\newcommand{\alphapom}{\alpha_{_{I\!\!P}}}

\newcommand{\lsim}{\raisebox{-0.5mm}{$\stackrel{<}{\scriptstyle{\sim}}$}}
\newcommand{\gsim}{\raisebox{-0.5mm}{$\stackrel{>}{\scriptstyle{\sim}}$}}

\begin{document}
\begin{titlepage}{GLAS-PPE/98--03}{27$^{\underline{\rm{th}}}$ June 1998}
\title{Highlights and Open Questions from ZEUS}
\author{Anthony T. Doyle\thanks{
Alexander von Humboldt fellow (Hamburg II University), 
supported by DESY and PPARC.
}\\
on behalf of the ZEUS Collaboration.}
\begin{abstract}
The latest results from the ZEUS experiment, as presented
at the DIS98 workshop, are
reviewed. A brief introduction to the status of the experiment is given.
The review focuses on three areas: multi-scale problems in QCD including 
results on jets, charm, the dynamics of $F_2$ at low $Q^2$, forward jets
and fragmentation
functions; diffractive structure including results on $\Upsilon$
production, inclusive diffraction in DIS, event shapes and leading baryon
production; and, the measurement and interpretation of the high-$Q^2$ 
neutral and charged current cross-sections.

\vspace{0.5cm}
\centerline{\em Introductory talk presented at the DIS98 Workshop,}
\centerline{\em Brussels, April 1998.}
\vspace{0.5cm}
\centerline{\em Slides are available from}
\centerline{\em http://www-zeus.desy.de/conferences98/\#dis98}
\end{abstract}
\newpage
\end{titlepage}

{\noindent\it HERA Progress:}
Since last year's DIS workshop, 
the data available for analysis at ZEUS has more than doubled,
corresponding to an integrated luminosity 
from the 1994-97 data-taking periods of 46.6~pb$^{-1}$.
This large data sample presents a significant challenge to the 
experimentalists in providing precise data in the many areas which 
HERA can uniquely access.
The understanding of the ZEUS detector is being improved
using the large data samples to calibrate in situ.
To illustrate the current level of understanding 
of the detector: 
the integrated luminosity is currently known to better than $\pm 1.5\%$;
the overall momentum scale of the central tracking detector has been 
established at below the 0.3\% level
following calibration using elastic $J/\psi$ data;
the electron energy scale in the barrel calorimeter has been 
calibrated to $\pm1\%$ using DIS data; and,
the hadronic energy scale has been determined within $\pm 3\%$ using
DIS data and cross-checked using dijet events.
Developments in the calibration of the detector, combined with the
improved statistics, enable increasingly precise as well as new measurements
to be performed, as discussed
below. 

\section{Multi-Scale Problems in QCD} 
At HERA there is always at least one soft QCD scale, parameterised in 
terms of the parton densities of the proton which require input from data. 
Quantities are measured with an additional hard scale, such as 
the photon virtuality, $Q^2$, in inclusive $F_2$ measurements.
The measurements are then compared to perturbative calculations
where the cross-sections may be factorised 
and the soft input constrained. 
In general, an additional soft (hadronisation) scale enters 
wherever part of the hadronic final state is measured.
In addition, in resolved photoproduction processes, the photon may also be 
described as a source of partons introducing an additional soft scale 
which again requires data input. 
An important point in the development of our understanding of QCD is to ensure
that the measurements are well-defined theoretically at the required level of
precision. 
These multi-scale measurements can then be confronted with 
the QCD predictions to test whether our current understanding 
is sufficient to describe the process under consideration.
A wide range of measurements can be made at HERA which test our understanding
of QCD and its factorisation properties.

{\it Photon Structure:~\cite{joost}}
Before discussing the latest preliminary measurements, 
it is worthwhile to 
note the uncertainties, as discussed in recently published 
results on inclusive jet production using the cone algorithm.~\cite{incjet} 
Experimentally the uncertainty on the jet energy scale of $\pm 3\%$
dominates the uncertainty on the cross-sections. 
Theoretically, the scale dependence 
on the renormalisation and factorisation
scales varied between $E_T^{jet}/4 < \mu_{F} = \mu_{R} < E_T^{jet}$,
is minimised for a cone radius $R \sim 0.6$, as observed earlier at 
the Tevatron.
The variation of $R_{sep}$, the two-jet merging parameter, leads to similar 
uncertainties on the cross-section for
$R < R_{sep} < 2R$.
In order to minimise the theoretical uncertainties
due to merging/seed-finding ambiguities, 
the iterative $k_T$-algorithm has been adopted.
An inclusive dijet analysis was performed requiring 
$E_T^{jet_1} > 14$~GeV and $E_T^{jet_2} > 11$~GeV.
\begin{figure}[hbt]
\vspace{-0.5cm}
\centerline{\psfig{file=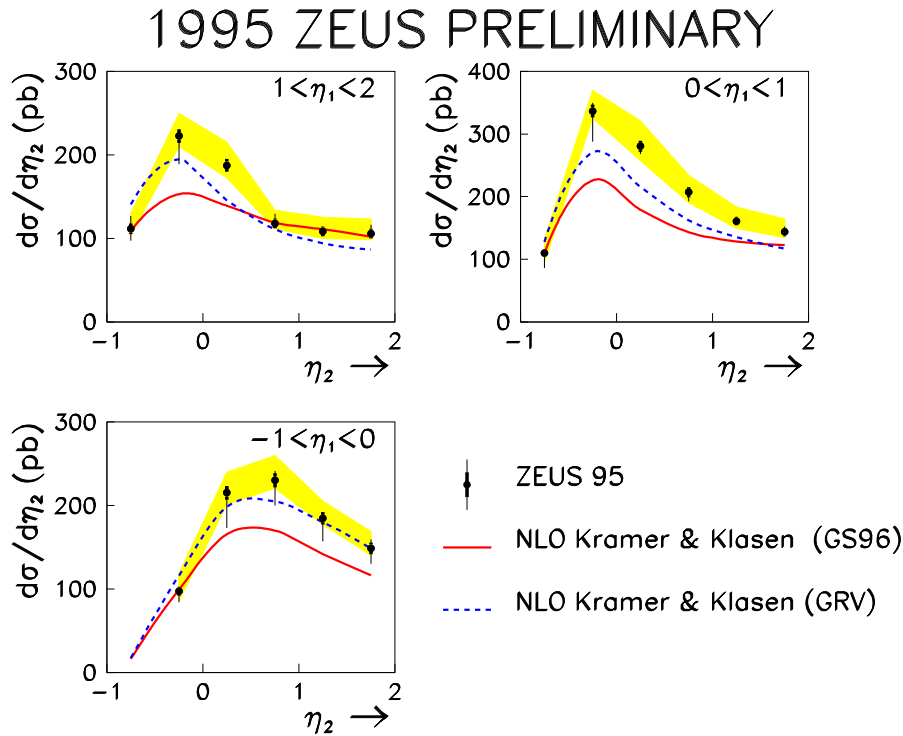,width=12cm}}
\vspace{-1cm}
\caption{\label{etah}
Inclusive dijet photoproduction cross-sections.}
\end{figure}
At these $E_T^{jet}$ values, comparisons with Monte Carlo indicate that
multiple interactions (not included in the NLO calculations)
are not required to describe the observed $x_\gamma$ distributions or the 
jet profiles.
To gain greater sensitivity to the internal structure of the photon
the measurements are made at high-$y$ ($0.5 < y < 0.85$), 
corresponding to the highest accessible photon energies.
In Fig.~\ref{etah}
the $d\sigma/d\eta_2$ distribution for the second-highest energy jet 
is presented for fixed intervals of $\eta_1$. 
The shaded band represents the hadronic energy scale uncertainty.
The hadronisation uncertainties have been estimated to be $\sim$ 10\%,
the scale uncertainties are estimated at $\sim$ 10\%
while the proton parton densities are well constrained
in the probed $x$ region ($x~\sim~10^{-2}$).
The data are thus sensitive to the choice of photon parton densities, 
as illustrated by the comparison of the the data with the 
GS96 and GRV parameterisations.
The ZEUS data has now reached a level of precision where the photon parton
densities can be discriminated: a global analysis of photon parton densities
incorporating such data is therefore required.

{\it Multijet Structure:~\cite{laurel}} 
In order to probe the QCD matrix elements at a deeper level, an inclusive
three-jet analysis has been performed using the $k_T$-algorithm 
for jets with $E_T^{jet} > 6$~GeV. The cross-section for such processes
can be written as:
$$
    d\sigma({\sf \gamma p \rightarrow jjj}) =
\sum\! \int\!\!\! \int\! dx_{\sf \gamma} dx_{\sf p}
\underbrace{f_{{\sf 1}\!/\!{\sf \gamma}}(x_{\sf \gamma})
f_{{\sf 2}\!/\!{\sf p}}(x_{\sf p})}_{\mbox{\normalsize$M^{\sf jjj}$}}
{\underbrace{|{\cal M}_{{\sf 12}\rightarrow
{\sf 345}}|^2}_{\mbox{\normalsize$\cos \theta_3$, $\psi_3$}}}
{\underbrace{d PS}_{\mbox{\normalsize $X_3$,$X_4$}}}
$$
\begin{figure}[htb]
\centerline{\psfig{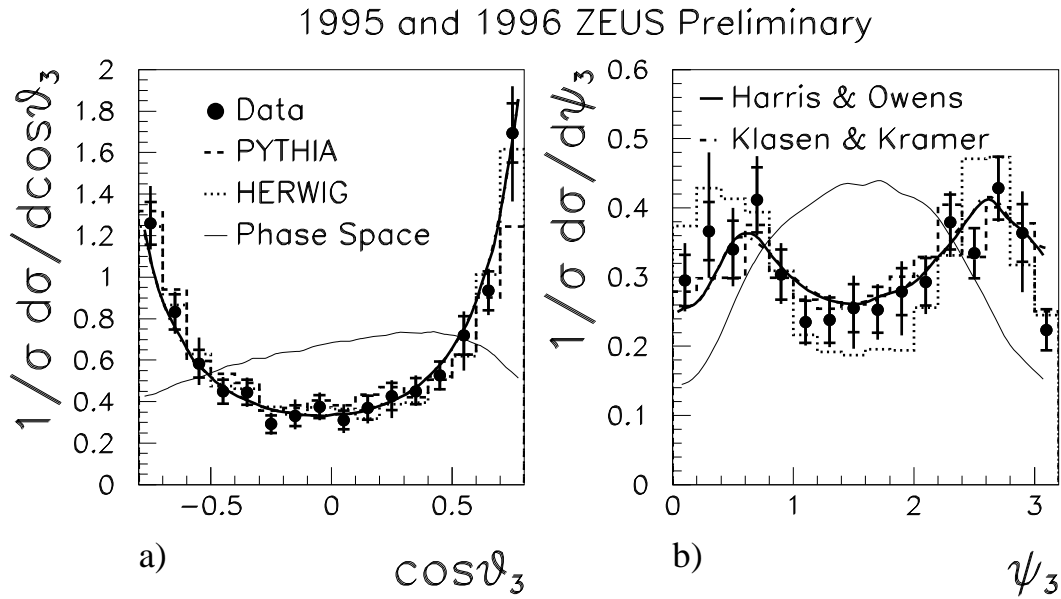}}
\vspace{-0.5cm}
\caption{\label{mudy}
Angular distributions for inclusive three-jet events compared to Monte Carlo
simulations
denoted by the histograms and the ${\cal{O}}(\alpha\alpha_S^2)$ calculations
discussed in the text.}
\end{figure}

\noindent 
Here, the measured invariant mass distribution, $M^{\sf jjj}$, is governed
by the photon and proton parton densities, $f_{{\sf 1}\!/\!{\sf \gamma}}(x_{\sf \gamma})$ and 
$f_{{\sf 2}\!/\!{\sf p}}(x_{\sf p})$,
whereas the scaled energies of the jets, $X_3$ and $X_4$, are controlled 
largely by phase space.
The measured angular distributions in the three-jet centre of mass 
are $\cos \theta_3$, where $\theta_3$ is the angle the highest-energy jet 
makes w.r.t. the beam axis,
and $\psi_3$, the angle of the three-jet plane w.r.t. the beam axis.
The $\cos \theta_3$ distribution is determined by the spin of the primary 
exchange and the distribution of
$\psi_3$ is related to the coherence property of the radiated (lowest-energy)
jet to lie in the plane of the beam and the highest-energy jet.
These are
shown in Fig.~\ref{mudy}, compared to ${\cal{O}}(\alpha\alpha_S^2)$
calculations (thick line) as well as to a phase space calculation (thin 
line)
where the spin of the partons is ignored. 
The comparisons constitute a refined test of the photoproduction
QCD matrix elements
${\cal M}_{{\sf 12}\rightarrow {\sf 345}}$.

\begin{figure}[htb]
\rightline{{ZEUS 1996-97 preliminary}\hspace{2cm}}
\vspace{-1.5cm}
  \centering
\mbox{
\subfigure[$D^*$ $p_\perp$ cross-section.]
{\psfig{file=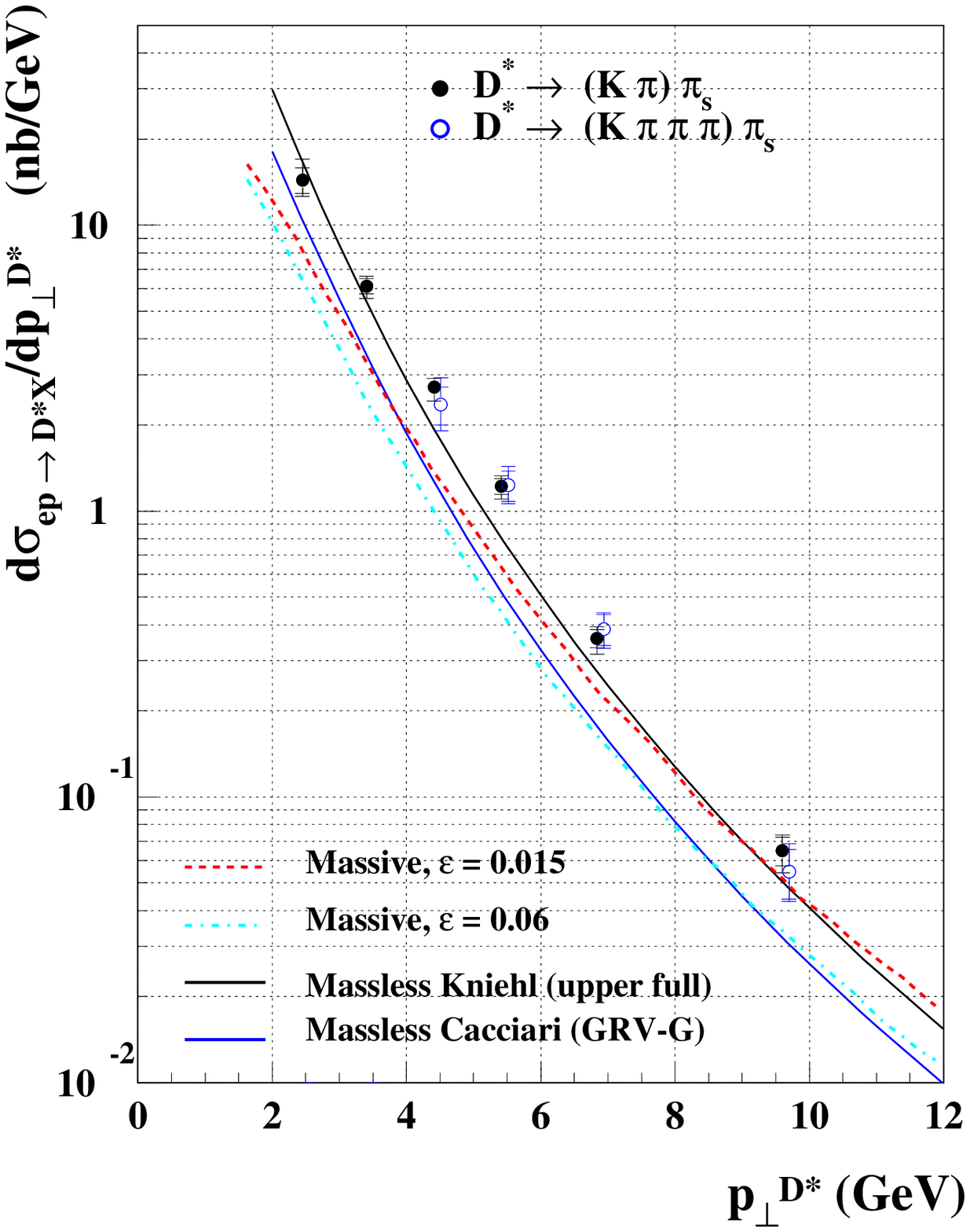,width=.45\textwidth}}\quad
\subfigure[$d\sigma/d\eta^{D^*}$ for $p_\perp > 3$~GeV.]
{\psfig{file=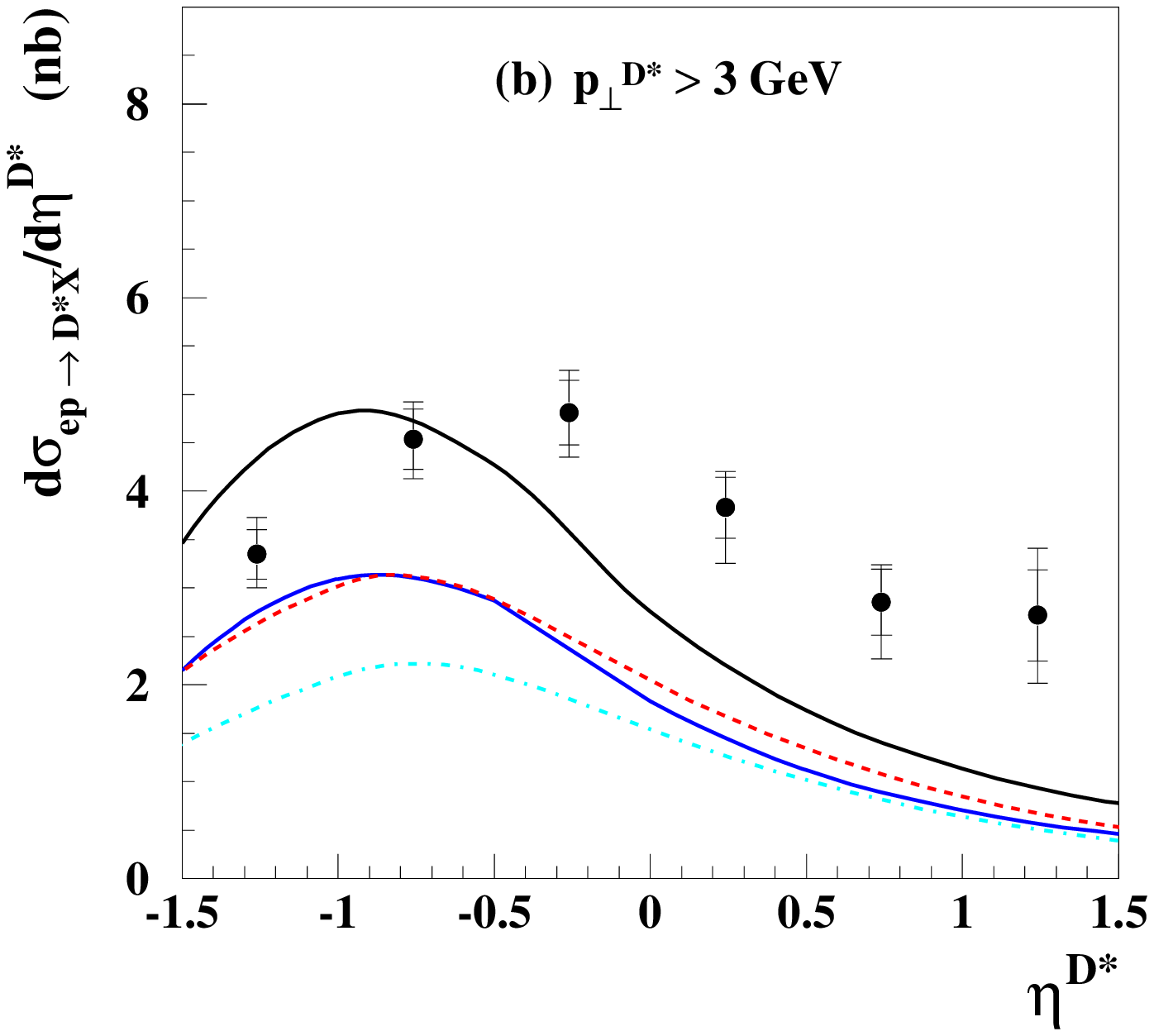,width=.5\textwidth}}
}
  \caption[]{\label{dstr}
$D^*$ $ep$ cross-sections compared to NLO
calculations.}
\end{figure}

{\it Charm in Photoproduction:~\cite{mark}} 
Heavy flavour production introduces a new scale with which to test 
perturbative QCD.
Measurements of the $D^*\rightarrow (K \pi) \pi_s$
and $D^*\rightarrow (K \pi \pi \pi) \pi_s$ channels 
are shown to be in good agreement in Fig.~\ref{dstr}(a).
\noindent
Two approaches have been taken in the calculations: 
the ``massive" approach (dashed lines) where charm is
generated dynamically and divergences are regulated by the charm mass;
and the ``massless" approach (full lines) 
where charm production is activated at the 
$m_c$ threshold and massless approximations are used.
Uncertainties arise due to the choice of the effective charm mass,
and renormalisation and factorisation scales (here 
$\mu_R=\sqrt{m_c^2 + p_\perp^2}=\mu_F/2$) 
as well as the hardness of the charm decay to $D^*$,
characterised in terms of $\epsilon_c$ in the Peterson fragmentation
function for the ``massive" calculations.
The measured cross-section is typically underestimated in the calculations.
In Fig.~\ref{dstr}(b) this excess is observed to be predominantly in the 
forward direction. 
The open question is whether the data are more sensitive to the choice of
input photon structure function or to the $D^*$ fragmentation dynamics
as we approach the proton fragmentation region or, perhaps, anomalously large
contributions from $b\rightarrow c$ decays.
\begin{figure}
\centerline{\small{ZEUS 1996-97 preliminary}}
\centerline{\psfig{file=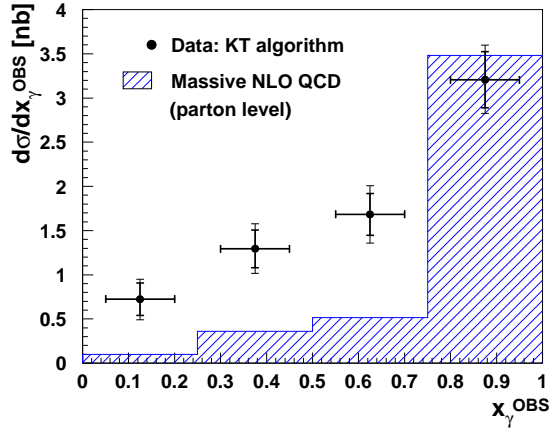,width=8cm}}
\caption{\label{dsxg}
$D^*$+dijet cross-section versus $x_\gamma^{OBS}$}
\end{figure}
\noindent Further information is provided by the measurement of 
associated dijets with 
$E_T^{jet_1} > 7$~GeV and $E_T^{jet_2} > 6$~GeV
shown in Fig.~\ref{dsxg}.
Here the measurement of $x_\gamma^{OBS}$ 
has a relatively large contribution 
at low-$x_\gamma$ whereas
the ``massive" NLO calculation is significantly more peaked towards one,
underestimating the low-$x_\gamma$ part of the cross-section.
The question here is whether 
the dynamical generation of charm in photoproduction is sufficient,
assuming that jet hadronisation effects, estimated at $\sim 10\%$ using Monte
Carlo simulations, are relatively less important.

\begin{figure}[p]
\centerline{
\hspace*{-33mm}
\epsfig{file=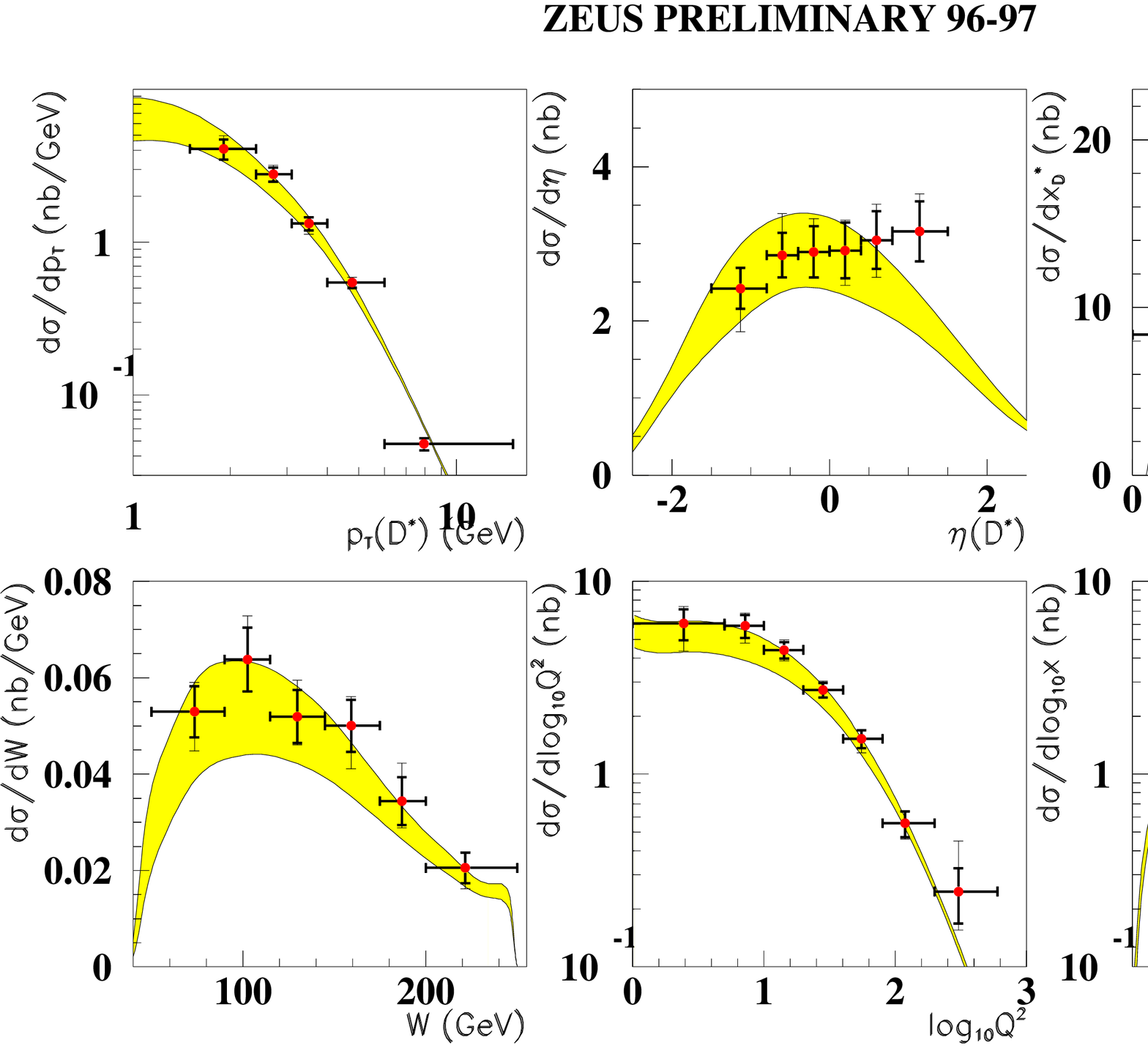,
bbllx=0pt,bblly=0,bburx=595,bbury=542,
width=10cm,clip=}
}
\vspace{-0.5cm}
\caption{\label{f2cd}
DIS $D^*$ cross-sections compared to NLO calculations (shaded band).}
\end{figure}

\begin{figure}[p]
\centerline{\psfig{file=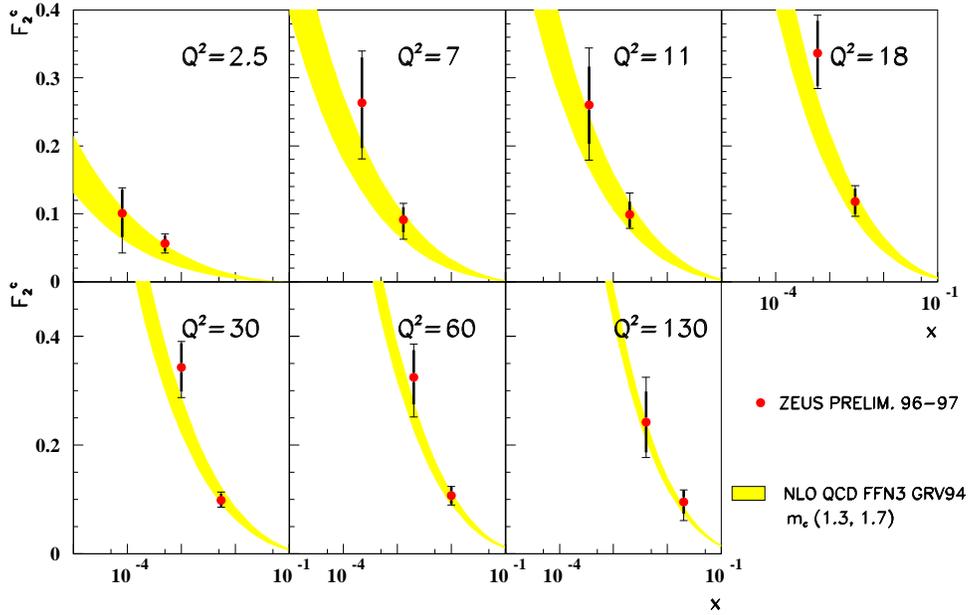,
bbllx=0pt,bblly=0,bburx=566,bbury=380,
width=13cm}}
\vspace{-0.5cm}
\caption{\label{f2ch}
$F_2^c(x, Q^2)$ as function of $x$ for fixed $Q^2$ compared to NLO 
calculations.}
\end{figure}

{\it Charm in DIS:~\cite{dave}} 
$D^*\rightarrow (K \pi) \pi_s$ measurements in DIS provide a
significant test of the gluon density of the proton 
determined from the scaling violations of $F_2$.
They will also help to constrain theoretical uncertainties in the
fits to $F_2$ where different prescriptions for heavy flavour effects
are adopted.
Compared to the photoproduction case, they remove the uncertainty 
due to the choice of photon PDFs 
and hence reduce the number of open question posed above.
The preliminary cross-section 
$\sigma^{ep\rightarrow D^*X} = 8.55 \pm 0.40 ^{+0.30}_{-0.24}$~nb
is measured in the range 
$1<Q^2<600$~GeV$^2$, $0.02<y<0.7,
1.5 < p_T(D^*) < 15$~GeV, and $\eta(D^*) < 1.5$.
In Fig.~\ref{f2cd}, the upper plots show the measurements 
of the hadronic final state variables $p_T(D^*)$, $\eta(D^*)$ and $x_{D^*}$,
the fractional momentum the $D^*$ in the $\gamma^* p$ rest frame:
the data agree with the massive NLO calculations where 
$\epsilon_c = 0.035$, except perhaps at lower
$x_{D^*}$ corresponding to higher $\eta(D^*)$.
In addition the kinematic variables, $W$, $Q^2$ and $x$ shown in the 
lower plots are in good agreement with the NLO calculations:
it is therefore reasonable to extrapolate the measured cross-section 
to the full $\{\eta(D^*), p_T(D^*) \}$ range 
\renewcommand{\thefootnote}{1}
\footnote{This procedure neglects the
possibility of additional contributions outside the measured region due, for
example, to intrinsic charm.}
to determine $F_2^c(x, Q^2)$ via the expression\\
$$\frac{d^2\sigma_{c\bar{c}X}}{dx\,dQ^2} = 
\frac{2 \pi \alpha^2}{x Q^4} 
[ (1 + (1-y)^2) F_2^c(x, Q^2) - y^2 F_L^c(x, Q^2)].$$
\noindent In Fig.~\ref{f2ch}, the $F_2^c(x, Q^2)$ data 
mirror the rise of $F_2$ at small $x$. 
The data are in agreement with the GRV94 PDFs, where the band represents 
an estimated theoretical  
uncertainty due to the effective charm mass ($m_c = 1.5\pm 0.2$~GeV).
This comparison verifies the steep rise of the gluon 
density at low $x$ with a precision of $\simeq 15-20\%$. 

\newpage
\begin{figure}[htb]
  \centering
\mbox{
\subfigure[$\lambda_{\rm eff}$ versus $Q^2$.]
{\psfig{file=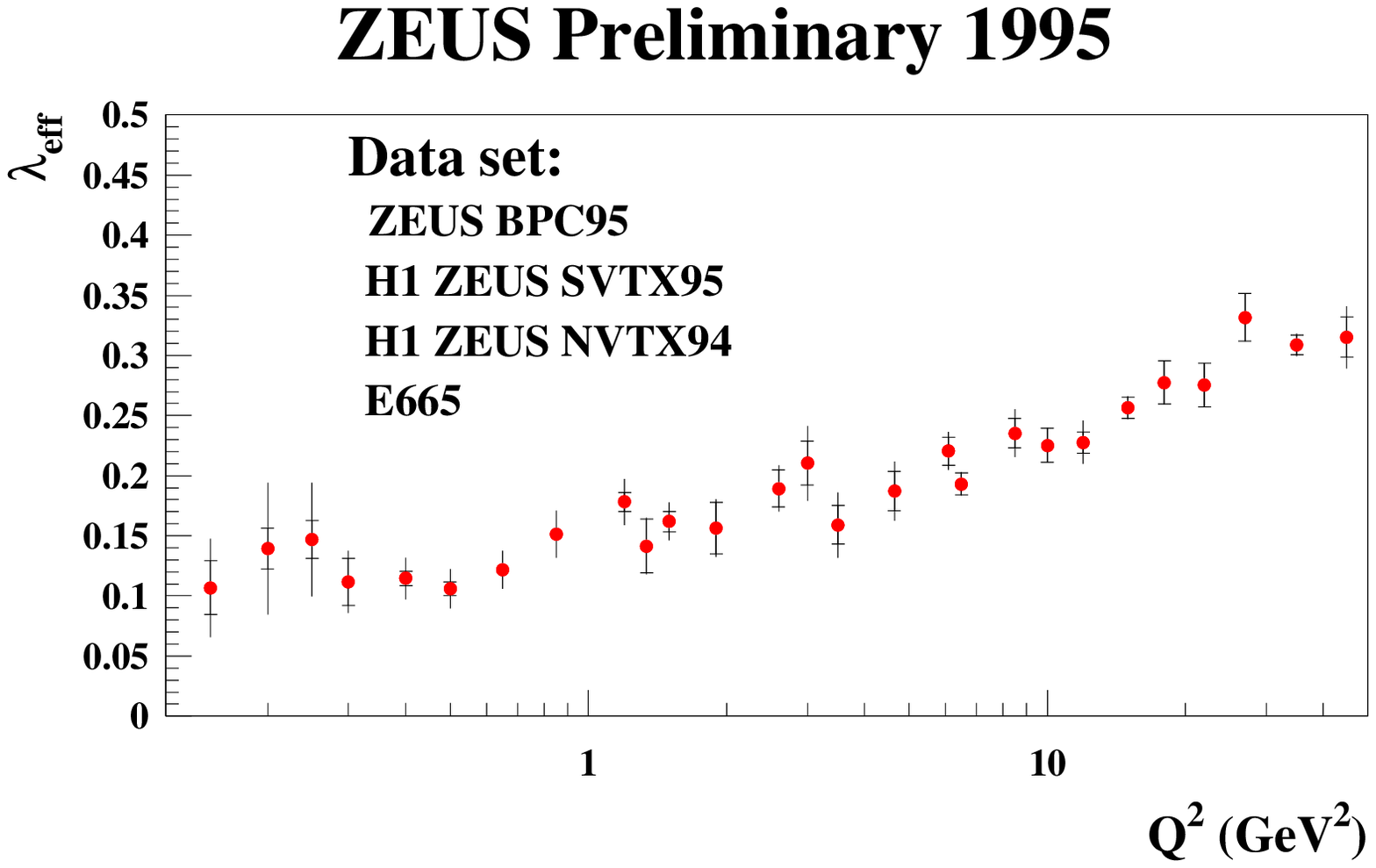,
bbllx=70pt,bblly=390,bburx=510,bbury=675
,width=.45\textwidth,height=.50\textwidth}}
\quad
\subfigure[$dF_2/dlog Q^2$ versus $x$.]
{\psfig{file=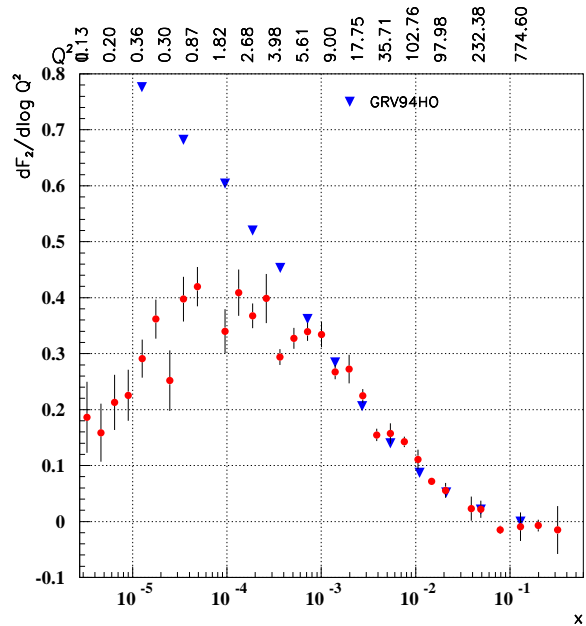,
bbllx=53pt,bblly=174,bburx=506,bbury=681
,width=.45\textwidth}}
}
  \caption[]{\label{xn02}
Fits to $F_2$ data ($\bullet$) exploring the transition region.}
\end{figure}

{\it Transition Region:~\cite{bernd}} 
The corresponding 
rise of $F_2$ with decreasing $x$, or equivalently
the rise of $\sigma^{tot}_{\gamma^{\star} p}$ with increasing $W$,  
has stimulated significant theoretical developments
in the understanding of QCD at high energies. One challenge is to
explore how and where the transition occurs from soft to hard physics
and interpret low-$Q^2$ data. 
In order to relate the low-$Q^2$ and
$Q^2~=~0$ data, 
a GVMD (Generalised Vector Meson Dominance) analysis has been performed.
This analysis relates the 
virtual photon cross-section to the real cross-section via
$\sigma^{tot}_{\gamma^{\star} p} = \sigma^{tot}_{\gamma p}
\cdot M^2_0 / (M^2_0 + Q^2)$,
for fixed $W$ ($\sigma_L$ contributions at small $Q^2$
lead to a small correction of $\sigma^{tot}$).
A good description of the ZEUS BPC data measured in the range 
$0.1<Q^2<0.65$~GeV$^2$ is found with
$M^2_0 = 0.53 \pm 0.04 \pm 0.10$~GeV$^2$.
Extrapolating to $Q^2=0$~GeV$^2$, the corresponding 
$W^{2(\alphapom(0)-1)}$ dependence
is given by the pomeron intercept value\\
\centerline{
$\alphapom(0)_{BPC}$ = 1.145 $\pm$ 0.02($stat$) $\pm$ 0.04($sys$)
(preliminary)
}
to be compared with the Donnachie-Landshoff value $\alphapom(0) = 1.08$.
In this $Q^2$ range, the rise of the cross-section is therefore relatively
modest. This behaviour is also seen in the lower $Q^2$ points of
Fig.~\ref{xn02}(a). 
Here additional datsets are incorporated in fits to the $F_2$ data
of the form
$F_2 = c\cdot x^{\lambda_{\rm eff}}\,|_{Q^2}$.
The parameter $\lambda_{\rm eff} \simeq \alphapom(0) - 1$ for 
$x~<~0.01$ is 
then plotted as a function of $Q^2$.
A relatively slow transition from $\lambda_{\rm eff} \simeq 0.1$ 
is observed with increasing $Q^2$. 
This rise of $F_2$ with decreasing $x$ is intimately coupled to the
scaling violations via the gluon density 
(in leading order $dF_2/dlog Q^2 \sim xg(x)$
neglecting sea quark contributions).
In Fig.~\ref{xn02}(b), fits of the form 
$F_2 = a + b \cdot log(Q^2)|_x$ have been performed to the published 
HERA
data and the pre-HERA prediction from the GRV94 PDFs.~\cite{allen} 
For $x~\lsim~10^{-4}$, corresponding to $<\! Q^2\! > \lsim~2$~GeV$^2$
there is a qualitative change in behaviour where the scaling violations
stabilise and then decrease for lower-$x$ values, a behaviour which is 
not reproduced by the GRV94 PDFs.
\begin{floatingfigure}[r]{60mm}
\centerline{\psfig{file=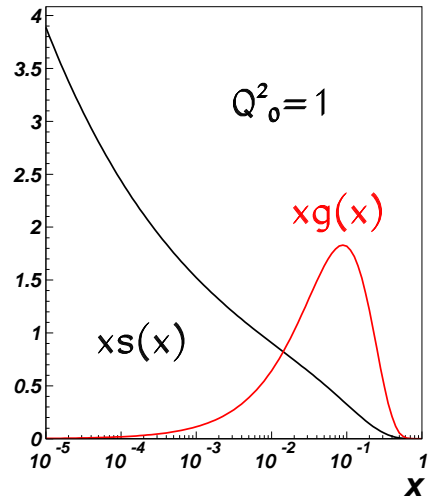,
bbllx=60pt,bblly=100,bburx=540,bbury=668,width=55mm}}
\vspace{-0.5cm}
\caption{\label{xn03}
Low $Q^2$ parton densities.}
\end{floatingfigure}
\noindent 
The question is whether this scaling violation behaviour and the slow onset of
the rise of $F_2$ with decreasing $x$ can be simultaneously
understood. 
A DGLAP NLO fit to the 
$Q^2 > 1$~GeV$^2$ data (not shown) describes the data, 
demonstrating that there is sufficient flexibility in such an 
approach to go down to relatively low $Q^2$.
However, the relatively stable scaling violations observed 
around $<\! Q^2\! > \sim 2$~GeV$^2$ in 
Fig.~\ref{xn02}(b) yield a gluon contribution which 
is rapidly diminishing
at small-$x$ and which is significantly smaller than the sea quark 
contribution for 
small starting scales, as illustrated in 
Fig.~\ref{xn03}:
in this low $Q^2$ region the sea appears to be driving the gluon at low-$x$. 
For larger $Q^2$ values the gluon dominates the sea and 
we have an intuitively appealing picture where gluons radiate sea quarks.
Whether such low-$Q^2$ partons 
are universally valid could be tested using e.g. low-$Q^2$ $F_2^c$
data.

\begin{figure}
\centerline{\psfig{file=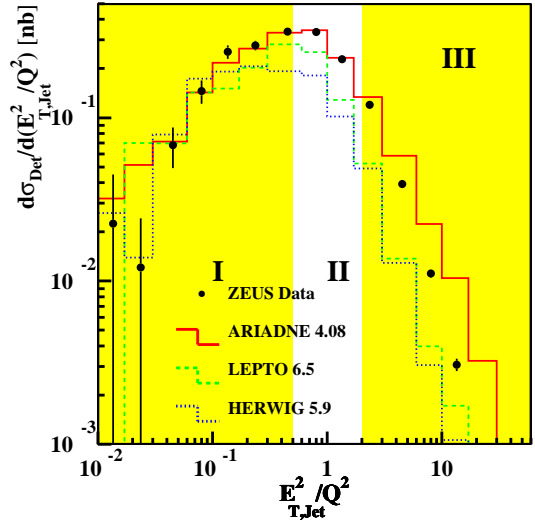,width=8cm}}
\vspace{-0.5cm}
\caption{\label{fjet}
Observed $E_{T}^{2}/Q^{2}$ distribution.}
\end{figure}
\begin{figure}[hbt]
\centerline{\psfig{file=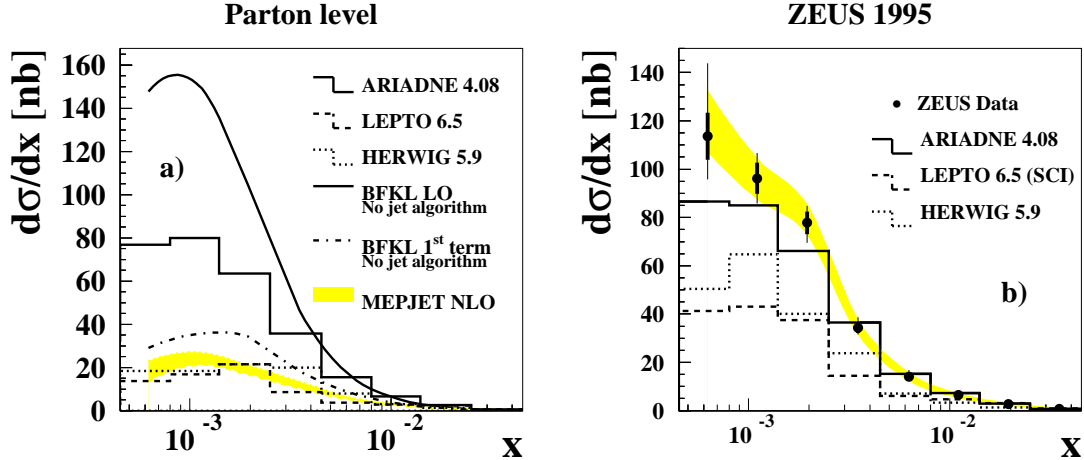,
width=14cm,
bbllx=50pt,bblly=450pt,bburx=600pt,bbury=720pt,clip=}}
\vspace{-0.5cm}
\caption{\label{fjxb}Forward jet production cross-sections at 
(a) parton level and (b) hadron level. The data are compared to 
various predictions discussed in the text.}
\end{figure}

{\it Forward Jet Production:~\cite{michael}} 
Why does $F_2$ rise?
In the DGLAP approach the $x$ dependence is
an input determined at a starting scale $Q_o^2$
and evolved in $Q^2$. In the BFKL approach the $x$-dependence
has recently been calculated in NLO.
The underlying dynamics may be tested using semi-inclusive
forward jet measurements in low-$x$ events.
Jets with $E_T^2 \sim Q^2$ and $x_{jet} \gsim x$, where 
$x_{jet}$ is the momentum fraction of the jet relative to the incoming proton,
are selected in order to enhance 
BFKL-like contributions where forward gluons may be emitted at relatively 
large $E_T$.
In Fig.~\ref{fjet} the jets observed at detector level are shown
as a function of $E_T^2/Q^2$ compared to three Monte Carlo simulations:
LEPTO~6.5 and HERWIG~5.9 are DGLAP-based models such that gluons emitted at
successively larger $x_{gluon} \sim x_{jet}$ 
have successively lower $E_T$ whereas 
the colour dipole model ARIADNE~4.08 
incorporates a BFKL feature that gluons are not strongly ordered.
\noindent Three regions are identified: 
I - $E_T^2 < Q^2/2$, the "DGLAP" region, where all models
approximately describe the data;
II - $Q^2/2 < E_T^2 < 2Q^2$, the "BFKL" region where the DGLAP-based models
(LEPTO~6.5 and HERWIG~5.9) fall below the data; and,
III - $E_T^2 > 2Q^2$, where all models fail 
and we may need to describe the DIS data in terms of a virtual
photon whose structure is being resolved by the high-$E_T$ jets.
The cross-section is evaluated in region II 
as a function of $x$ for $x_{jet} > 0.036$ and $E_T > 5$~GeV 
in Fig.~\ref{fjxb}. 
BFKL dynamics leads to an enhancement of the forward jet
production cross-section proportional to $(x_{jet}/x)^{\alphapom -1}$ 
over the ${\cal O}(\alpha\alpha_S^2)$ calculation.
\noindent
As shown in Fig.~\ref{fjxb}(a) there is a significant difference between 
the ${\cal{O}}(\alpha\alpha_S^2)$ MEPJET calculation (represented 
by the shaded band which includes 
the uncertainty on the renormalisation scale) compared to the 
leading-order BFKL prediction (full curve) at parton level. 
There are residual uncertainties in determining the hadron-to-parton
level corrections and therefore the measurement in Fig.~\ref{fjxb}(b)
is presented at hadron level.
The rise of $F_2$ at small $x$ is mirrored by the rise of the
measured forward jet cross-section which is not described by 
the DGLAP-based models. A consistent description of the $F_2$ and
forward-jet data represents a considerable challenge to our understanding
of QCD.

{\it Fragmentation Functions:}
Recently published results on jet shapes 
have shown that the observed patterns of QCD radiation in high-$Q^2$ 
neutral current and charged current are similar to those observed in
$e^+e^-$ experiments.~\cite{mario}
\begin{figure}[hbt]
  \centering
\mbox{
\subfigure[$\ln(1/x_p)_{max}$ versus $\ln(Q)$.]
{\psfig{file=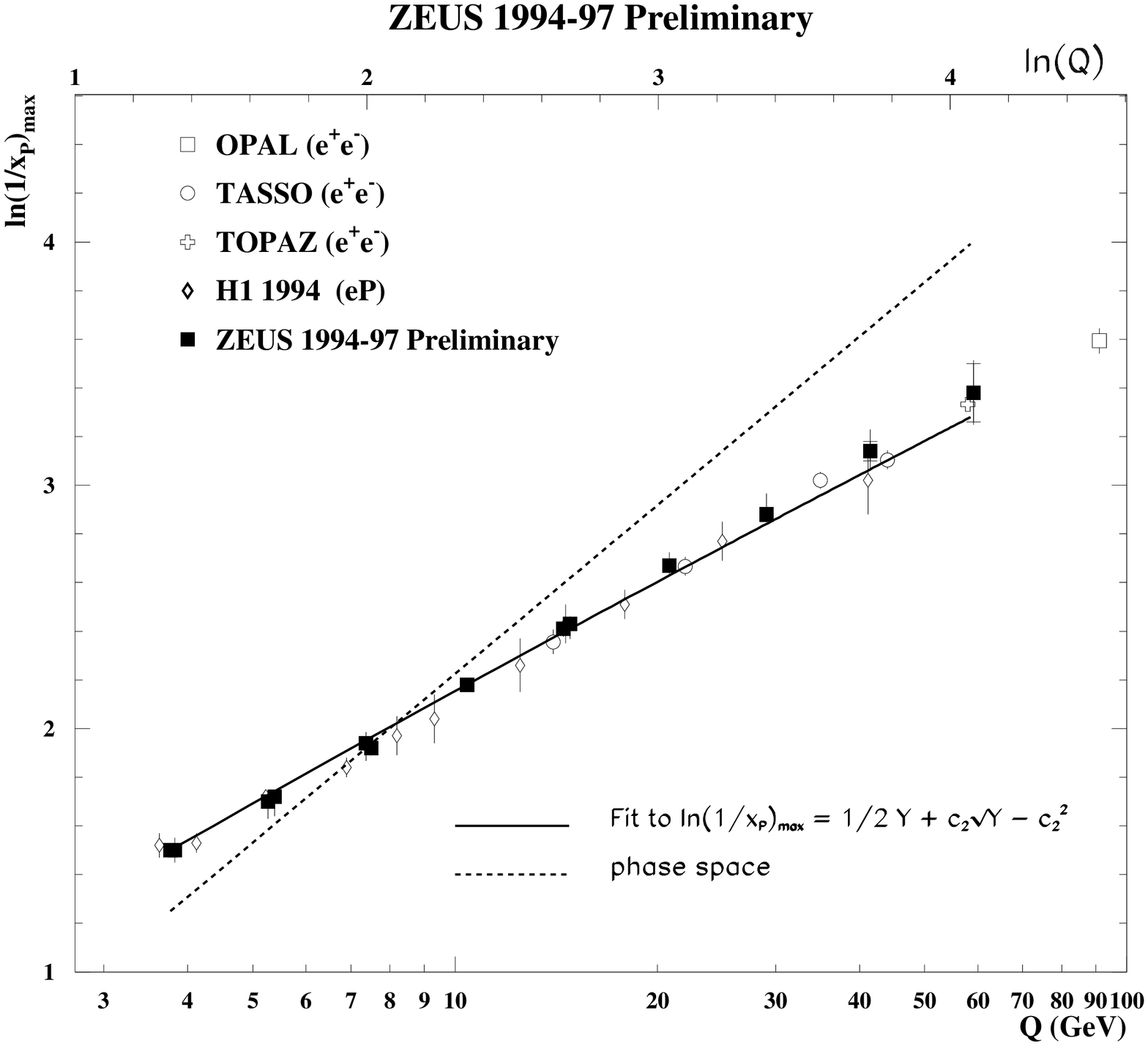,
height=.5\textwidth,width=.45\textwidth}}
\subfigure[$1/\sigma d\sigma/dx_p$ versus $Q^2$.]
{\psfig{file=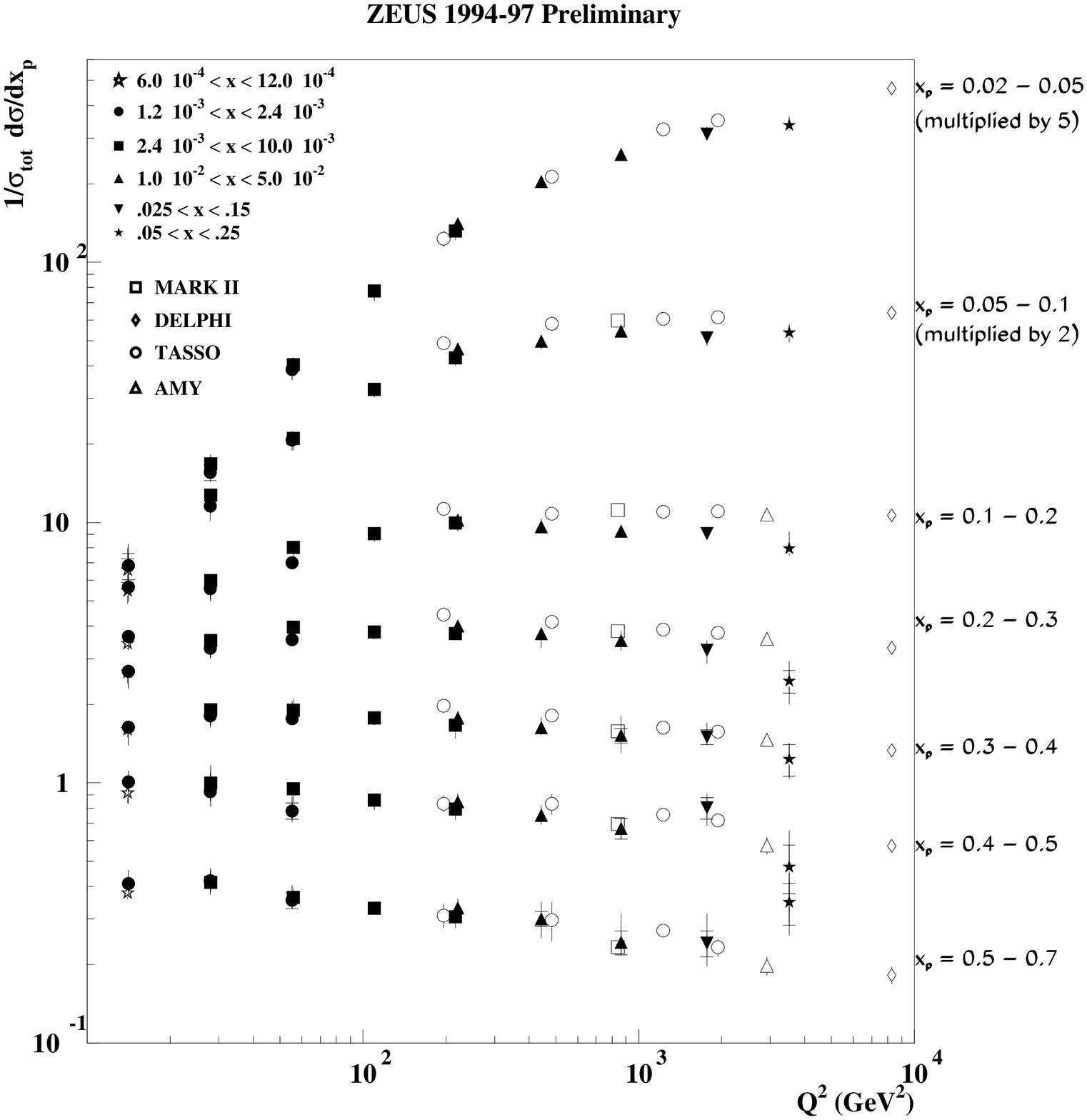,width=.5\textwidth}}
}
  \caption[]{Charged hadron fragmentation 
in the current region of the Breit frame.}
\label{frag}
\end{figure}
\noindent 
At HERA we are able to study these properties as a function of $Q^2$
in a single experiment and hence provide detailed information on quark
fragmentation properties.~\cite{jan}
These properties can be studied in the semi-soft limit by measuring 
the $\ln(1/x_p)$ distributions, where $x_p = 2p/Q$ is the scaled momentum 
of charged hadrons 
in the current region of the Breit frame. The observed Gaussian distributions
are then fitted within $\pm 1$ of the mean to yield the
$\ln(1/x_p)_{max}$ values as a function of $Q$ given in 
Fig.~\ref{frag}(a).
The precise data are consistent with data from $e^+e^-$ experiments 
establishing the universality of fragmentation over a large range of 
$Q$.
An MLLA+LPHD fit (indicated by the full line) 
of the form $\ln(1/x_p)_{max} = 1/2Y +c_2\sqrt(Y)-c_2^2$, where 
$Y = \ln Q/2\Lambda_{eff}$ and $c_2 = 0.52$ for three active flavours
in the cascading process, provides a reasonable description of the data with
$\Lambda_{eff} \simeq 245$~MeV.
The high-statistics data enables the region of high $x_p$ to be studied.
In Fig.~\ref{frag}(b) the fragmentation function data are presented 
for different $x_p$ intervals as a function of $Q^2$
(in various ranges of $x$). 
At high $x_p$ and higher $Q^2$ the fragmentation 
functions exhibit negative scaling violations, consistent with a dominant 
QCD Compton process (c.f. high-$x$ structure function data).
The DIS data (full symbols) 
are observed to be reasonably consistent with $e^+e^-$ data (open symbols), 
but systematically lower at intermediate $x_p$ values.
Comparisons with NLO calculations where fragmentation functions extracted from
$e^+e^-$ data are implemented can describe the DIS data 
(see Fig.~2 in~\cite{jan}).
However care needs to be taken to explicitly include 
strange-quark fragmentation 
functions which are systematically softer than those of the up/down quarks.
This is important
since the production of strange quarks from the proton sea in DIS is 
suppressed (by a factor $\simeq 0.2$) compared to those from $e^+e^-$
annihilation.
%
In order to compare with recent $1/Q^2$
power-correction calculations, 
the fragmentation function data were also presented as function of
$x_\| = 2 p\cdot q /Q^2 = 2p_Z/Q$, where the $Z$ direction is defined by 
the virtual-photon proton axis.
\begin{figure}
\centerline{\psfig{file=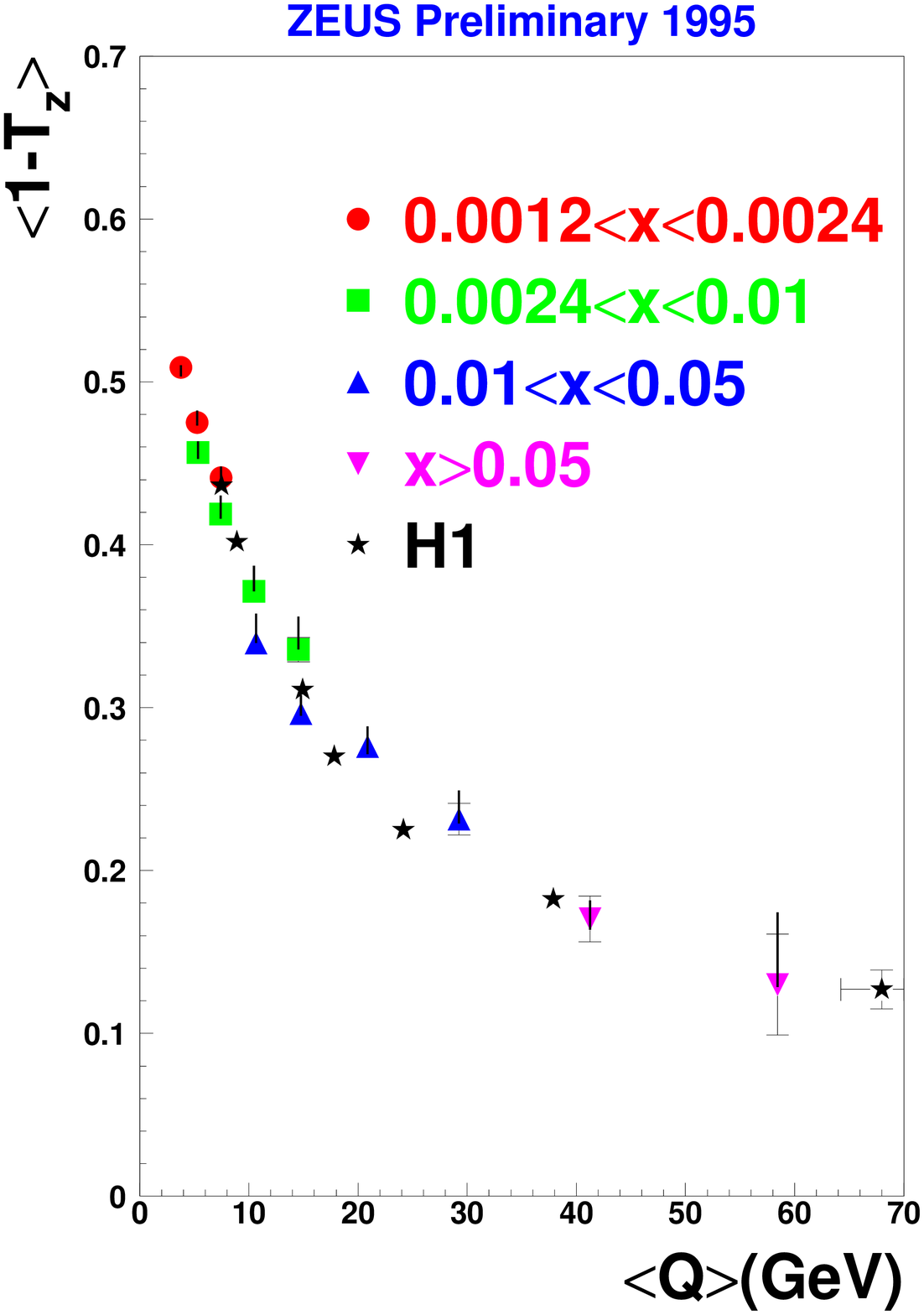,width=6.5cm}}
\vspace{-0.5cm}
\caption{\label{evsh}
$<\! 1-T_Z\! >$ 
versus $Q$.}
\end{figure}
\noindent 
Another approach to investigate
the r\^ole of such power corrections
is to sum over these momenta and 
measure the corresponding thrust distributions 
$T_Z = 2 \Sigma p_Z/ \Sigma p$. 
A series of 
event shape variables have been measured 
in the current region of the Breit frame
where the power corrections, 
determined from renormalon calculations,
are expected to be 
characterised by a universal $\overline{\alpha_{o}}$ and 
to fall as $1/Q$. 
In Fig.~\ref{evsh} the 
measurements 
using charged tracks in the current region of the Breit frame 
are displayed as $<\! 1-T_Z \!>$ versus $Q$.~\cite{rob}   
The characteristic behaviour is
in reasonable agreement with published H1 results given at slightly different 
$x$~values. 
Other event shape
variables are chosen which are relatively 
insensitive, in varying degrees, to soft gluon emission
and collinear parton branchings in order to determine whether
a universal $\overline{\alpha_{o}}$ can be applied.
As a first step, comparisons with NLO calculations 
illustrate the need for additional power correction 
terms (see Fig.~2 in~\cite{rob}).
\section{Diffraction} 
The study of diffractive phenomena and their relation to the inclusive
cross-section at HERA may hold the key to understanding how a 
``pomeron" structure emerges from QCD.
What we have learned so far is that a single soft pomeron does not describe 
$all$ diffractive data measured at HERA.
As the photon virtuality, $Q^2$, or the vector meson mass increases
the cross-sections rise rapidly with increasing $W^2$.
As we investigate the pomeron more closely, a new type of dynamical pomeron
may begin to play a r\^ole whose structure is being
measured in DIS.
These inclusive diffractive 
data are consistent with a partonic description of the exchanged object
which may be described by perturbative QCD.
At this workshop, new results were presented on 
$\phi$ electroproduction;~\cite{sergey}  
the $t$-dependence of $\rho$, $\phi$ and $J/\psi$ 
photoproduction;~\cite{teresa}  
$\psi'/J/\psi$ and $\Upsilon$ photoproduction;~\cite{alessia}   
the determination of $F_2^{D(2)}$ and the
diffractive/total cross-section in DIS;~\cite{henri}   
diffractive dijet photoproduction cross-sections;~\cite{didijet}  
event shapes in DIS using the LPS;~\cite{riko}   
and, leading baryon production in DIS.~\cite{alberto}  
These results will help to clarify our understanding of processes which 
represent a significant fraction, $\sim 10\%$, of the corresponding 
total cross-sections, 
yet cannot be understood in terms of a simple hadronisation mechanism.

\begin{figure}[htb]
\centerline{ZEUS 1995 preliminary}
  \centering
\mbox{
\subfigure[$\rho^0$ trajectory.]
{\psfig{file=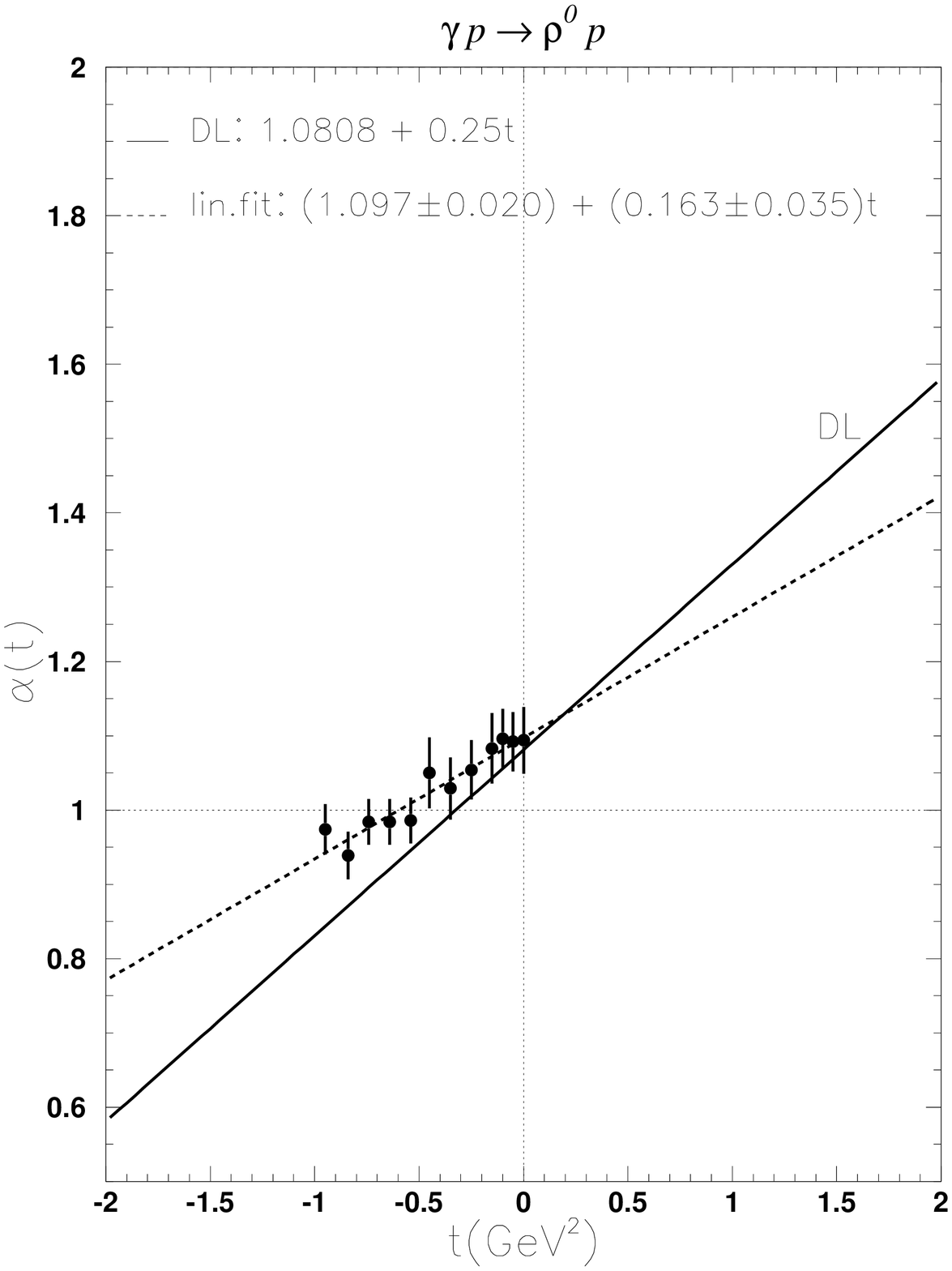,
bbllx=13pt,bblly=65,bburx=548,bbury=775,width=.45\textwidth}}\quad
\subfigure[$J/\psi$ trajectory.]
{\psfig{file=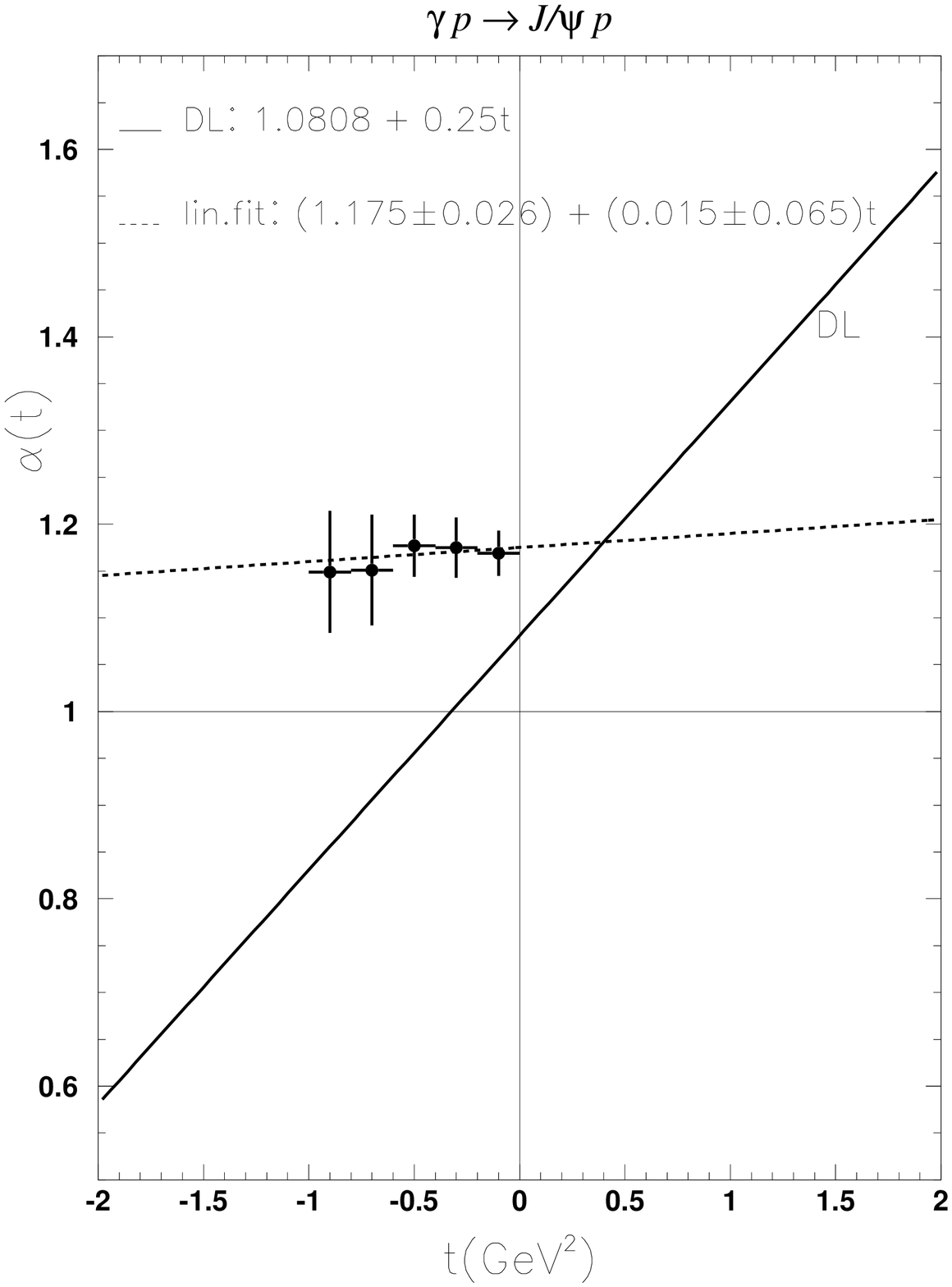,
bbllx=13pt,bblly=65,bburx=548,bbury=775,width=.45\textwidth}}
}
  \caption[]{\label{dalp}
Pomeron
trajectories determined from fits to HERA and lower $W$ data.}
\end{figure}

{\it Vector meson $t$ dependences:~\cite{teresa}} 
Diffraction is characterised
by a steeply-falling dependence 
of the cross-section as a function of $t$,
the momentum transferred at the proton vertex.
This characteristic fall-off increases approximately linearly 
with increasing $W$. 
This ``shrinkage"
behaviour is built into the Donnachie-Landshoff pomeron\\
\centerline{$\alpha(t) = \alpha(0) + \alpha^\prime\cdot t = 1.08 + 0.25 \cdot t.$}
Measurements of the quasi-elastic ($\gamma p \rightarrow V p$)
and proton-dissociative ($\gamma p \rightarrow V N$) production 
of vector mesons as a function of $t$ have been performed.
The $W$ dependence of the cross-section at fixed values of $t$
allows a direct determination of the exchanged trajectory. 
In order to gain sufficient range in $W$ and hence constrain $\alpha^\prime$,
the ZEUS preliminary results are 
combined with lower-$W$ elastic measurements and H1 published results.
In Fig.~\ref{dalp}(a) the $\rho^0$ trajectory is shown to be a strong
function of $t$ with \\
\centerline{$\alpha(t)_{\rho^0}  = (1.097\pm 0.020) + (0.163\pm0.035)\cdot t 
~~({\rm preliminary})$}
for the linear fit where the errors are the combined 
$({\rm stat}\oplus{\rm sys})$ errors from each experiment. 
Similarly, the $\phi$ trajectory (where non-leading
Regge trajectories are suppressed due to the Zweig rule) is determined as \\
\centerline{$\alpha(t)_\phi = (1.083\pm 0.010) + (0.180\pm0.027)\cdot t 
~~({\rm preliminary}).$}
These results are in contrast to the $J/\psi$ trajectory shown in 
Fig.~\ref{dalp}(b) where \\
\centerline{$\alpha(t)_{J/\psi} = (1.175\pm 0.026) + (-0.015\pm0.065)\cdot t 
~~({\rm preliminary}).$}
The result for $\alpha^\prime$ is consistent with no shrinkage of the cross-section,
which is qualitatively expected from perturbative QCD which allows 
only a slow (logarithmic) dependence of the trajectory as a function of $t$.
These results herald an era of study where the gluon-dominated QCD structure 
at the proton vertex is probed using a diffracted heavy flavour probe. 

\begin{figure}[htb]
\centerline{ZEUS 1995-97 preliminary}
  \centering
\mbox{
\subfigure[$M_{\mu^+ \mu^-}$ spectrum.]
{\psfig{file=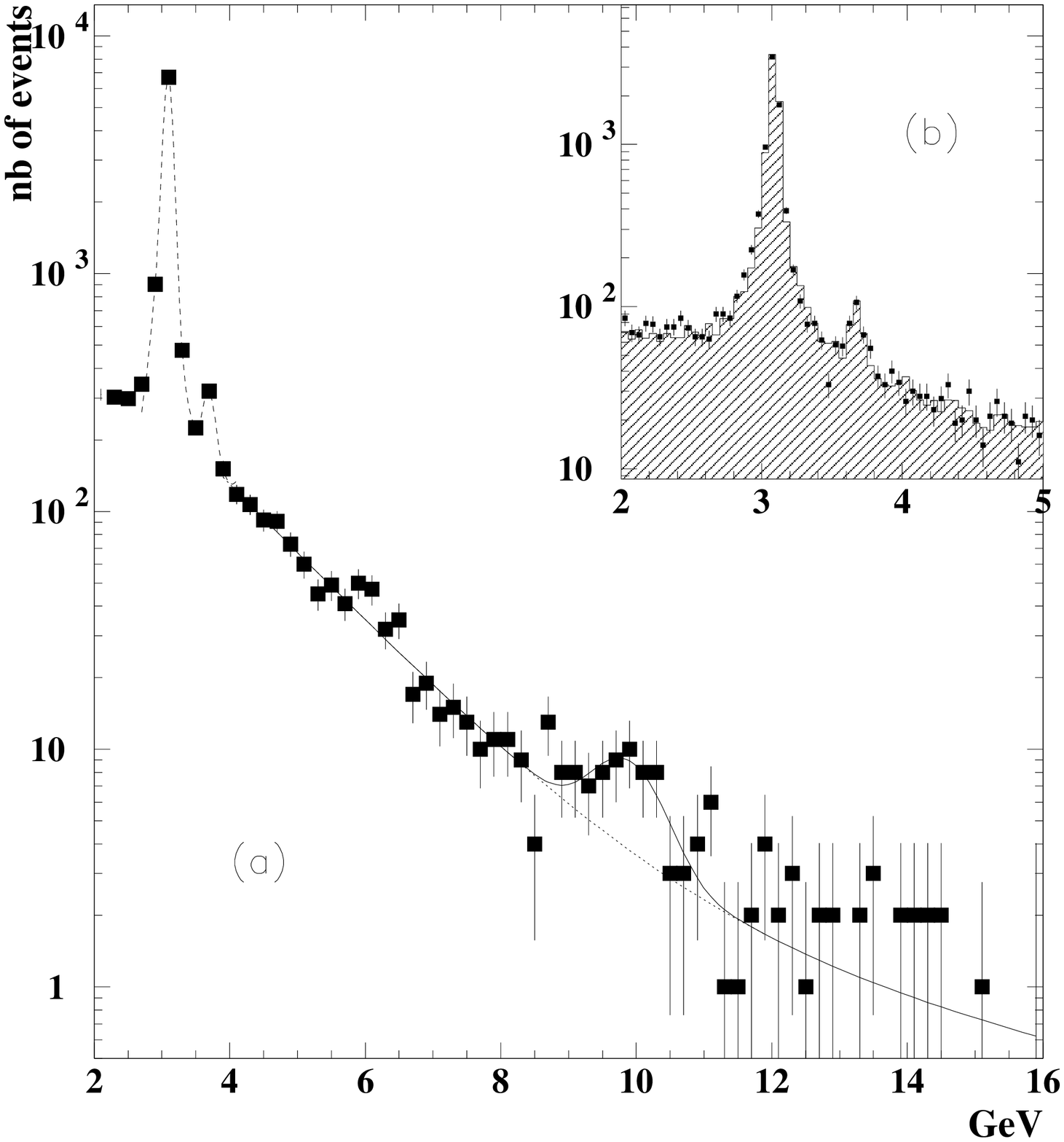,width=.40\textwidth}}\quad
\subfigure[ $\Upsilon$ cross-section versus $W$ compared to LO calculations.]
{\psfig{file=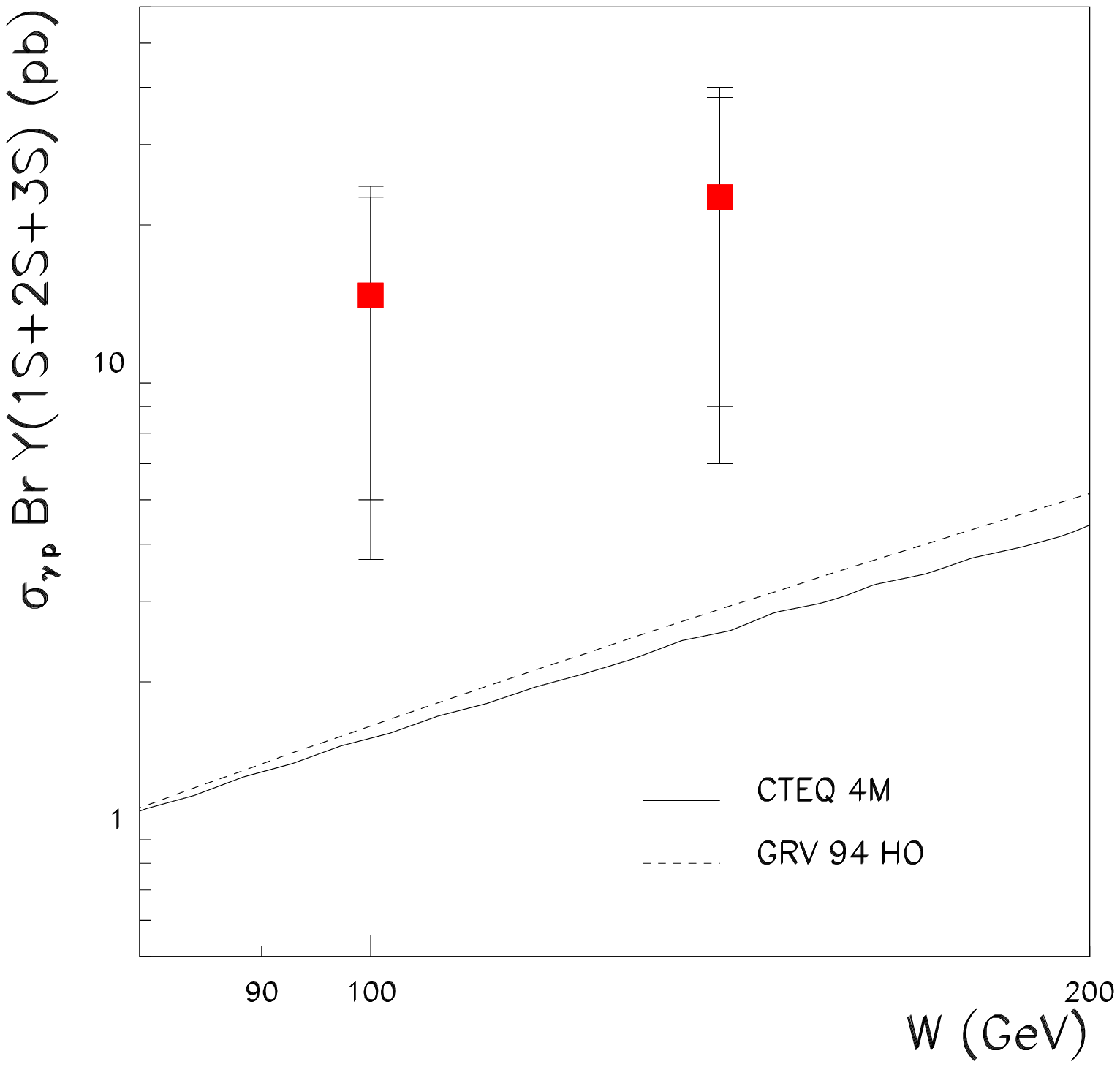,width=.50\textwidth}}
}
  \caption[]{\label{dups}
Measurement of the 
elastic $\Upsilon$ photoproduction cross-section.}
\end{figure}

{\it Higher Mass Vector Mesons:~\cite{alessia}}
Twenty years after the discovery of the
$\Upsilon$, the first observation of a signal in photoproduction 
is shown in Fig.~\ref{dups}(a) as a broad enhancement around
10 GeV in the di-muon mass spectrum.  The insert in 
Fig.~\ref{dups}(a) indicates the $J/\psi$ and $\psi'$ resonances:
the $\psi'/\psi$ production ratio is measured to be 
$R = 0.16 \pm 0.02 \pm 0.04$, where the largest
contribution to the systematic 
uncertainty is on the $\psi'\rightarrow \mu^+\mu^-$ branching ratio. 
The result is in agreement with
QCD calculations, determined by the wavefunction at the origin 
for the 1S and 2S states, of $\simeq 0.17$.
The observation of the $\Upsilon$ leads to the measurement of the elastic 
cross-section for the unresolved 
$\Upsilon(1S)$, $\Upsilon(2S)$ and $\Upsilon(3S)$
states shown in Fig.~\ref{dups}(b). 
Here the relative rates or muon production from the   
$\Upsilon$ states is determined from CDF data and applied as a correction to 
the $\Upsilon(1S)$ calculation. 
Comparison with the leading-order QCD calculations, 
where $\sigma_{\rm{diff}} \sim xg(x)^2 \otimes \hat\sigma $,
indicates that the measured cross-section is above these expectations.

\begin{figure}[p]
\centerline{\psfig{file=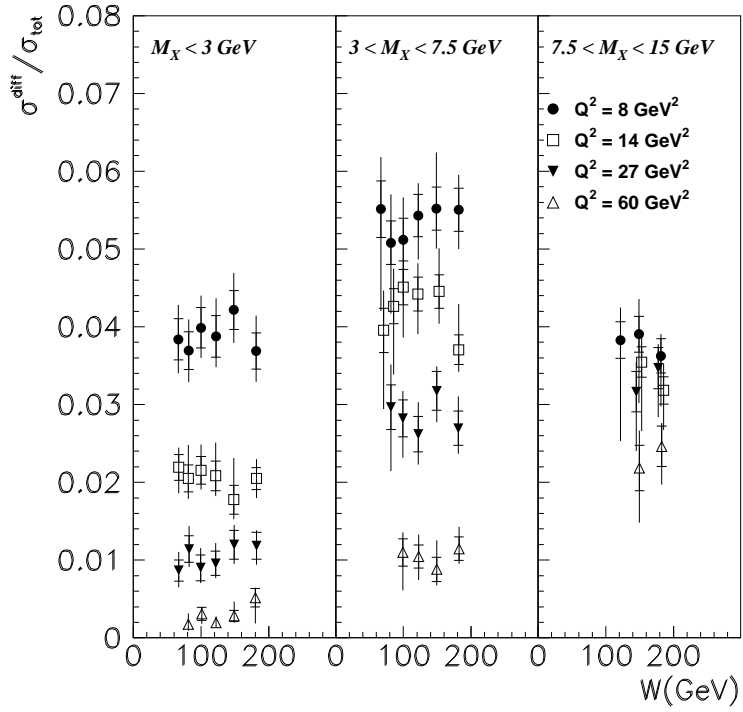,
bbllx=37pt,bblly=136,bburx=525,bbury=670,width=10cm}}
\vspace{-0.5cm}
\caption{\label{dif1}
Diffractive/total cross-section versus $W$ in various $M_X$ ranges.}
\end{figure}

{\it Diffractive Structure Functions:~\cite{henri}}
A new era for diffraction was opened by the observation of large rapidity
gap events in DIS and their subsequent analysis in terms of a diffractive
cross-section. The diffractive contribution is identified as a
non-exponentially suppressed contribution at small masses, $M_X$,
of the dissociating virtual photon system.
In Fig.~\ref{dif1} the ratio of this diffractive contribution to the 
total virtual photon proton cross-section is given as a function of $W$ 
for various $M_X$ intervals.
\noindent 
\begin{figure}[p]
\centerline{\psfig{file=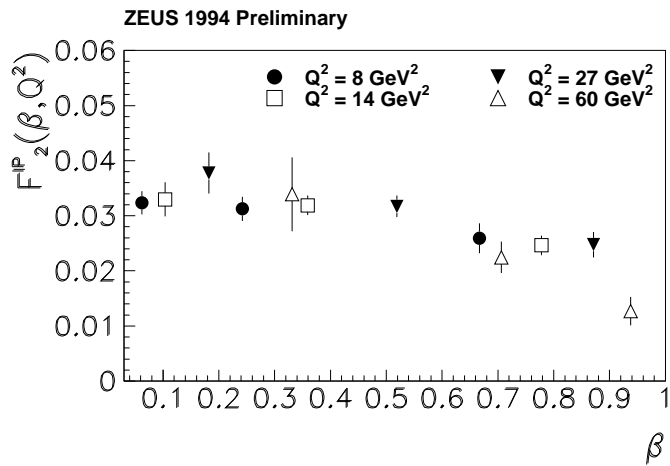,width=10cm}}
\vspace{-0.5cm}
\caption{\label{dif2}
$F_2^{D(2)}$ versus $\beta$.}
\end{figure}
\noindent 
An approximately constant ratio with $W$ indicates a diffractive contribution
which rises with a similar $W$ dependence.
This simple observation is contrary 
to the n\"{a}ive expectation where the diffractive contribution is 
identified with the forward part of the scattering amplitude and would 
therefore rise twice as quickly as the total cross-section as a function 
of $W$.
The rise of the diffractive cross-section with $W$ can be 
parameterised in terms of a power law, yielding 
$\alphapom(0) = 1.16 \pm 0.01 \pm 0.02$ (preliminary), after integration 
over $t$ with a mean exponential slope, $b=7$~GeV$^{-2}$ 
and assuming $\alpha^\prime = 0.25$~GeV$^{-2}$.
In order to understand the driving mechanism responsible for this rise,
the cross-sections at fixed $M_X$ and $W$ are 
plotted in terms of scaling variables.
Integrating over
$\xpom$, the momentum fraction of the pomeron within the proton,
leads to the $\beta$ dependence of $F_2^{D(2)} (\beta,Q^2)$ 
($\equiv F_2^{\pom}$ specified at $\xpom = 0.0042$)
shown in
Fig.~\ref{dif2} where 
$\beta$ is the momentum fraction of the struck quark within the pomeron.
An approximately flat dependence on $\beta$ is observed 
and an approximate scaling in $Q^2$ emerges from analysis of the data.
The decreasing fraction of events in each $M_X$ interval with increasing 
$Q^2$ observed in
Fig.~\ref{dif1} can thus be seen as due to integrating over a 
decreasing $\beta \simeq Q^2/(Q^2+M_X^2)$ region which is approximately
flat in $\beta$.

{\it Diffractive Event Shapes:~\cite{riko}}
The measurements of the structure function of
the pomeron constrain various models of diffraction. 
These models may be discriminated
using event shape variables 
which are directly sensitive to the underlying partonic
structure. Similarly, the data can be directly compared with 
$e^+e^-$ data where the underlying gluon 
Bremsstrahlung 
structure is well known.
Tagging a leading proton in the LPS allows a wide range of 
$M_X$ up to 25~GeV 
to be explored for $\xpom < 0.03$ and $Q^2 > 4$~GeV$^2$.
\begin{figure}[hbt]
\centerline{\psfig{file=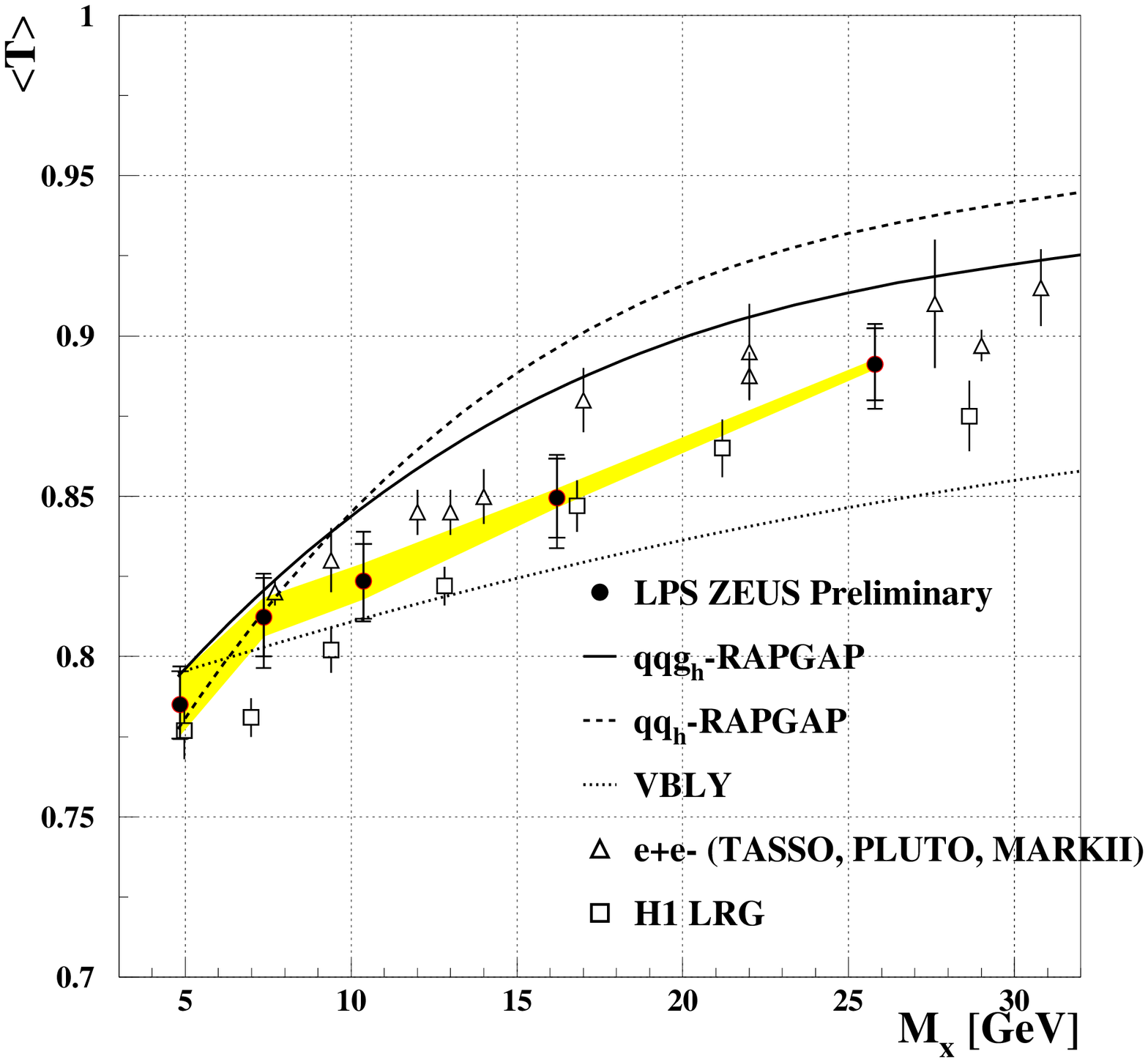,width=9cm}}
\vspace{-0.5cm}
\caption{\label{devs}  $<\!$Thrust$\!>$ versus $M_X$} 
\end{figure}
\noindent 
Measurements of the mean thrust are presented
as a function of $M_X$ in Fig.~\ref{devs}.
The LPS measurements are able to discriminate amongst various models
which assign different partonic structures to the pomeron.
The data exhibit reasonably 
similar 
values compared 
to $e^+e^-$ data for all values of $M_X$
suggesting that additional gluon contributions from the pomeron are relatively
small.
They are also consistent, but systematically higher than, 
the H1 data obtained using the large rapidity gap method.
Improved statistics, from existing data, 
will help to clarify whether the gluon contribution
of the pomeron determined from the scaling violations of
$F_2^{D(2)}$~\cite{henri} is 
consistent with that extracted from event shape variables~\cite{riko} 
and diffractive dijet photoproduction~\cite{didijet}.

{\it Leading Baryon Production:~\cite{alberto}}
Non-diffractive contributions play a r\^ole at higher values of $\xpom$. 
Measurements of forward proton production allow us to compare 
data with fragmentation models as well as models based on reggeon exchange.
Preliminary results for the corrected rates of proton production in the range
$0.60<x_L\equiv 1-\xpom <0.91$ for DIS at low and 
higher $Q^2$ ($0.1<Q^2<0.8$~GeV$^2$ and
$Q^2 > 4$~GeV$^2$) are ($13.0\pm0.5^{+0.7}_{-0.8}$)\% and
($12.7\pm0.3\pm0.9$)\%, respectively. 
The measured rates are typically higher than the
currently available fragmentation models by a factor of 1.5~to~2.
Measurements of uncorrected forward neutron production rates have 
been compared to reggeon exchange models: the comparisons indicate that 
the internal structure of the exchanged reggeon falls like $1-\beta$
at large $\beta$ (a dependence similar to $\pi$ (and higher) Regge exchanges), 
in contrast to the approximately flat behaviour with $\beta$ noted earlier 
for diffractive exchange.
The rates for neutron production are approximately the same for DIS,
photoproduction and even proton-gas interactions although the $W$ dependence 
of these cross-sections is significantly different. This 
suggests that ``diquark" fragmentation is a universal process 
which is to a large extent independent of the type of interaction
with the incident proton.

\section{High-$Q^2$ Cross-Sections}
The HERA collider provides a unique window to explore $ep$ interactions
at the highest energies, extending the range of momentum transfer $Q^2$
by about two orders of magnitude compared to fixed-target experiments.
As the HERA luminosity increases we explore the region of 
$Q^2 \sim 10^4$~GeV$^2$, where electroweak effects play a r\^ole. 
It is in this unexplored kinematic region that we are 
sensitive to deviations from the standard model (SM).
At last year's DIS97 workshop, H1 and ZEUS reported an excess of
events compared to the SM predictions from the
neutral current (NC) data taken during 1994 to 1996.
For the ZEUS analysis the observed rates agreed with expectations 
except for an excess at the highest $Q^2$ where two outstanding events 
with $Q^2 \simeq 40,000$~GeV$^2$ were observed from a luminosity of
20.1~pb$^{-1}$. 
\begin{figure}[htb]
\vspace{-0.5cm}
  \centering
\mbox{
\subfigure[Neutral Current Events.]
{\psfig{file=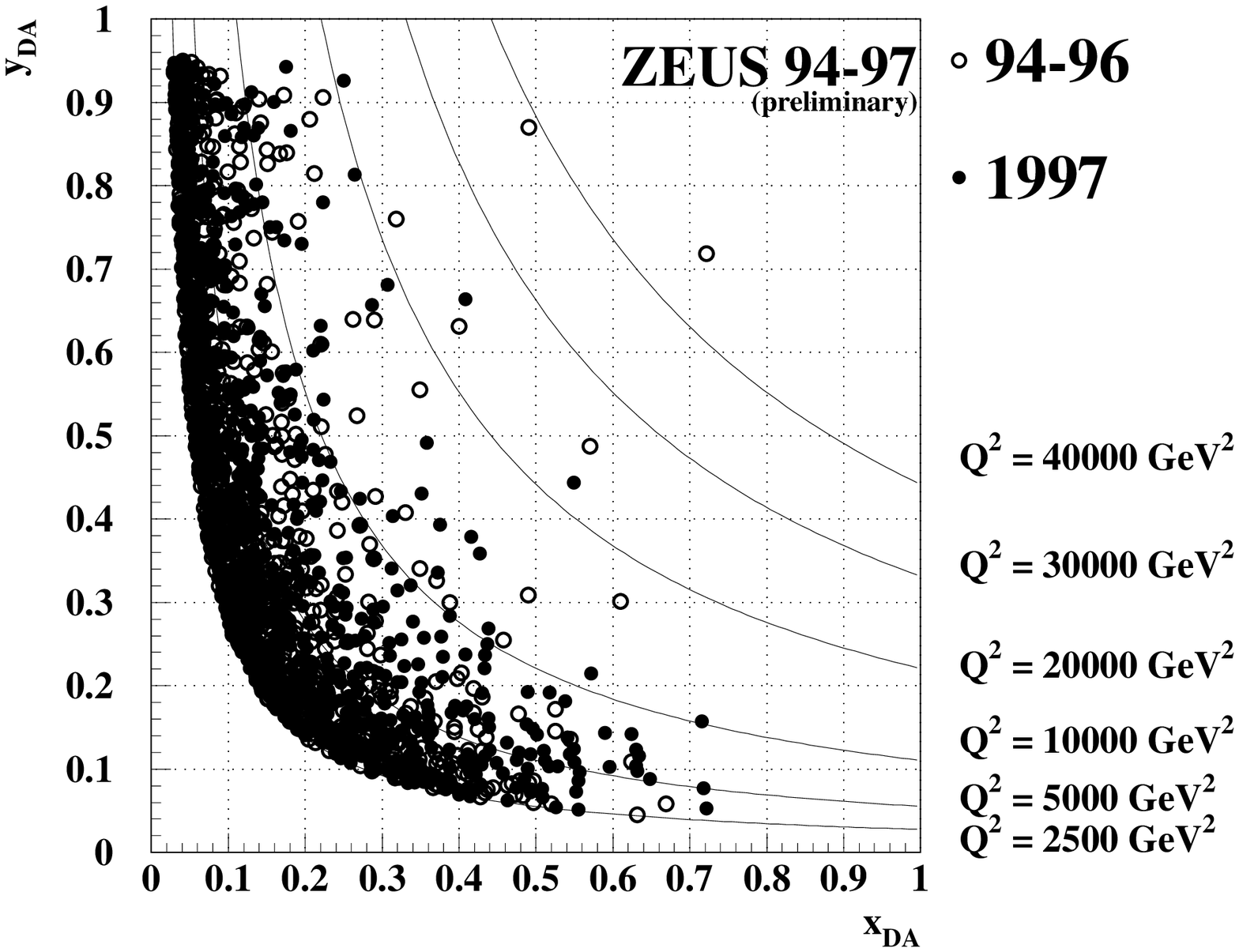,width=.58\textwidth}}
\subfigure[Charged Current Events.]
{\psfig{file=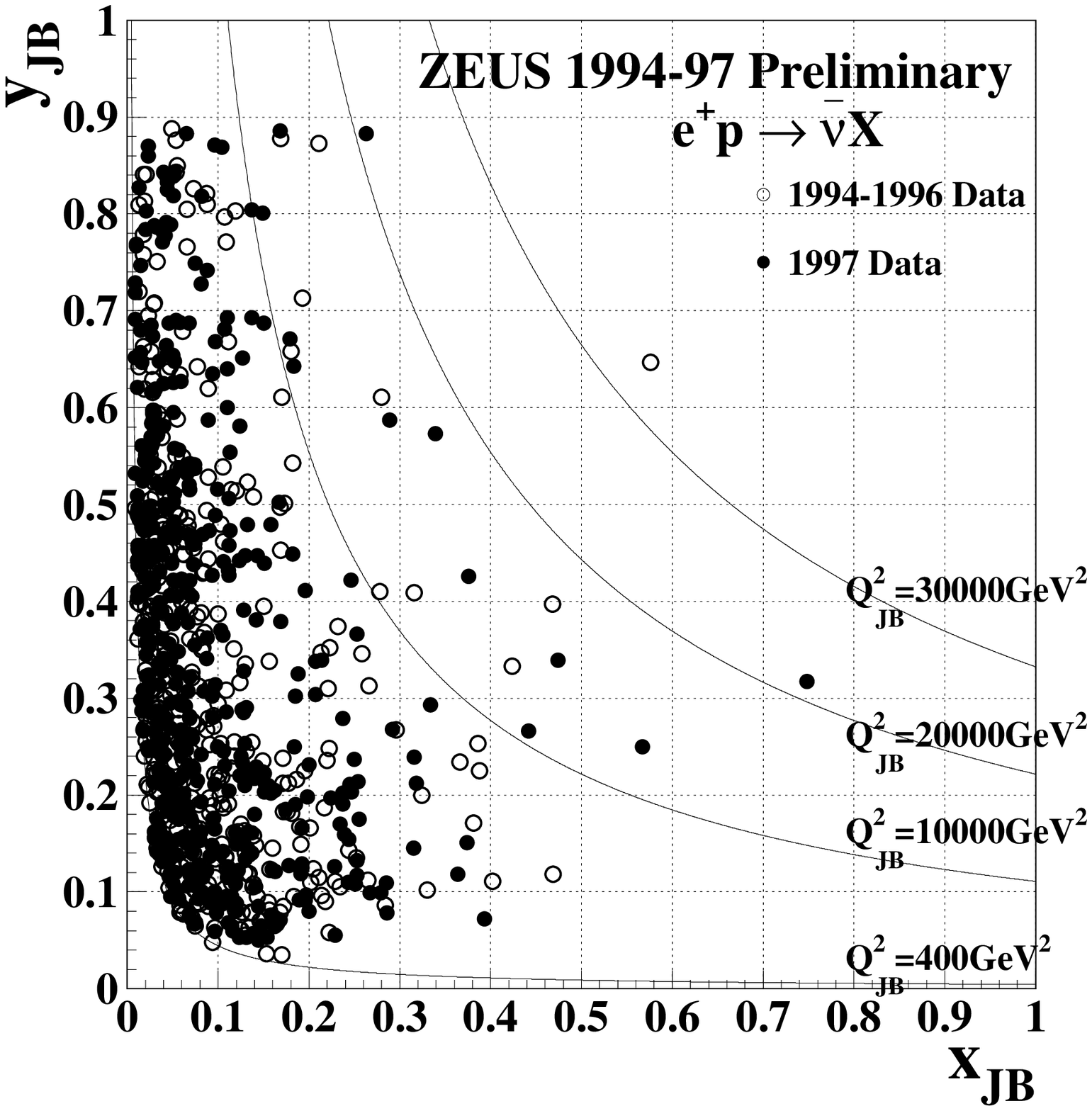,width=.45\textwidth}}
}
  \caption[]{\label{ccnc} Reconstructed $y$ versus $x$ scatter plots for (a) 
neutral current and (b) charged current events
from the 1994-96 ($\circ$) and 1997 ($\bullet$) data.}
\end{figure}
\noindent
In Fig.~\ref{ccnc}(a), these events clearly stand out but
no new NC outstanding 
events are observed in the 1997 data,
corresponding to a further 26.5~pb$^{-1}$ of data. 
In Fig.~\ref{ccnc}(b), one charged current (CC) event is observed at very high 
$Q^2 \simeq 30,000$~GeV$^2$ from the 1994-1996 data as well as 
two further events 
from 1997 at $Q^2 \simeq 20,000$~GeV$^2$. The number of events
is higher than expectations but is consistent with the standard model.
Attention has therefore focussed on measuring the cross-sections at the
highest accessible $Q^2$ values.
The theoretical uncertainty on the cross-sections 
was determined from a ZEUS QCD fit to
the structure function data on proton and deuteron targets from 
SLAC, BCDMS and NMC as well
as the neutrino measurements from CCFR, taking into account the correlation
amongst the 
systematic errors of each experiment. $\alpha_S(M^2_Z)$ was varied from 
0.113 to 0.123 and a 50\% uncertainty in the strange quark content was included.
In addition, various published PDFs with different models for charm evolution 
were used as well as fits incorporating E706 prompt photon data and CDF jet 
data. The results yielded SM cross-section uncertainties of 
$\simeq$~6-8\% on the NC cross-section and 
$\simeq$~6-12\% on the CC cross-section at 
the highest accessible $Q^2$ values.
These cross-sections therefore represent a benchmark for the standard model.
The cross-sections, discussed below, are corrected to the electroweak 
Born level 
and integrated over the complete $y$ range.

{\it Charged Current Cross-Sections:~\cite{gareth}}
Charged current events are identified by their missing transverse momentum
($p_T$) due to the escaping neutrino.
The cross-section is sensitive to the valence $d$-quark 
distribution in the proton:

$$\frac{d^2 \sigma_{e^+p}}{dx\, dQ^2} \simeq \frac{G_F^2}{2 \pi} 
\frac{1}{(1+Q^2/M_W^2)^2} [
\overline{u} + \overline{c} + (1-y)^2 (d + s)]. $$
$d\sigma^{CC} / dQ^2$ was measured for $Q^2 > 400$~GeV$^2$
using the Jacquet-Blondel method where $Q^2_{JB} = p_T^2/1-y_{JB}$,
with an RMS resolution on $Q^2$ of $\simeq 25\%$, reflecting the 
$35\%/\sqrt{E}$ hadronic energy resolution.
869 events were selected 
with backgrounds below $2.5\%$, diminishing with increasing $Q^2$.
The systematic uncertainties, mainly due to the hadronic energy scale
uncertainty of $\pm3\%$, correspond to $\sim 15\%$ uncertainties on the 
cross-section at lower $Q^2$ but increase at larger $Q^2$.
In the upper plot of Fig.~\ref{cc_xsec} the cross-section is observed
to fall over more than four orders of magnitude.
The ratio of the data to the SM, adopting the CTEQ4D PDF,
is shown in the lower plot of Fig.~\ref{cc_xsec} where
good agreement is observed up to $Q^2$ of $\simeq$~10,000~GeV$^2$. 
Comparison of the the data uncertainties 
with those from theory (shaded band) indicates that the data 
will help to
constrain the $d$-quark densities at large-$x$.
The cross-section is suppressed at lower
$Q^2$, due to the $1/(1+Q^2/M_W^2)^2$ propagator term: this characteristic
dependence on $Q^2$ has been fitted to yield a value for the mass of the
exchanged (space-like) $W$-boson of
\centerline{
$M_W = 78.6 ^{+ 2.5}_{- 2.4}(stat.) ^{+3.3}_{- 3.0}(syst.)\;\rm GeV$
(preliminary)}
with an additional PDF uncertainty of $\pm 1.5$~GeV.

Photoproduction of $W$-bosons decaying semi-leptonically 
has been investigated by searching for events with a high-$p_T$ lepton 
and missing $p_T$ with 46.6~pb$^{-1}$ of data. 
This is interesting in the context of the observed excess of
high-$p_T$ muons with associated missing $p_T$ 
observed by H1~\cite{tim}. 
In the ZEUS analysis four events are observed in the electron channel
where $2.22\pm0.02$ are expected from $W$ production and $1.24\pm0.35$
from various backgrounds. Similarly,
zero events are observed in the muon channel
where $0.46\pm0.02$ are expected from $W$ production and $0.84\pm0.23$
from other sources.
The ZEUS measurements enable 95\%CL limits to be set on 
$\sigma(W) (p_T^{miss} > 20$~GeV) of 2.5~pb and 2.0~pb in the electron
and muon channels, respectively.

\addtocounter{figure}{1}%
\begin{figure}[p]
  \includegraphics[angle=90,width=0.90\textwidth]{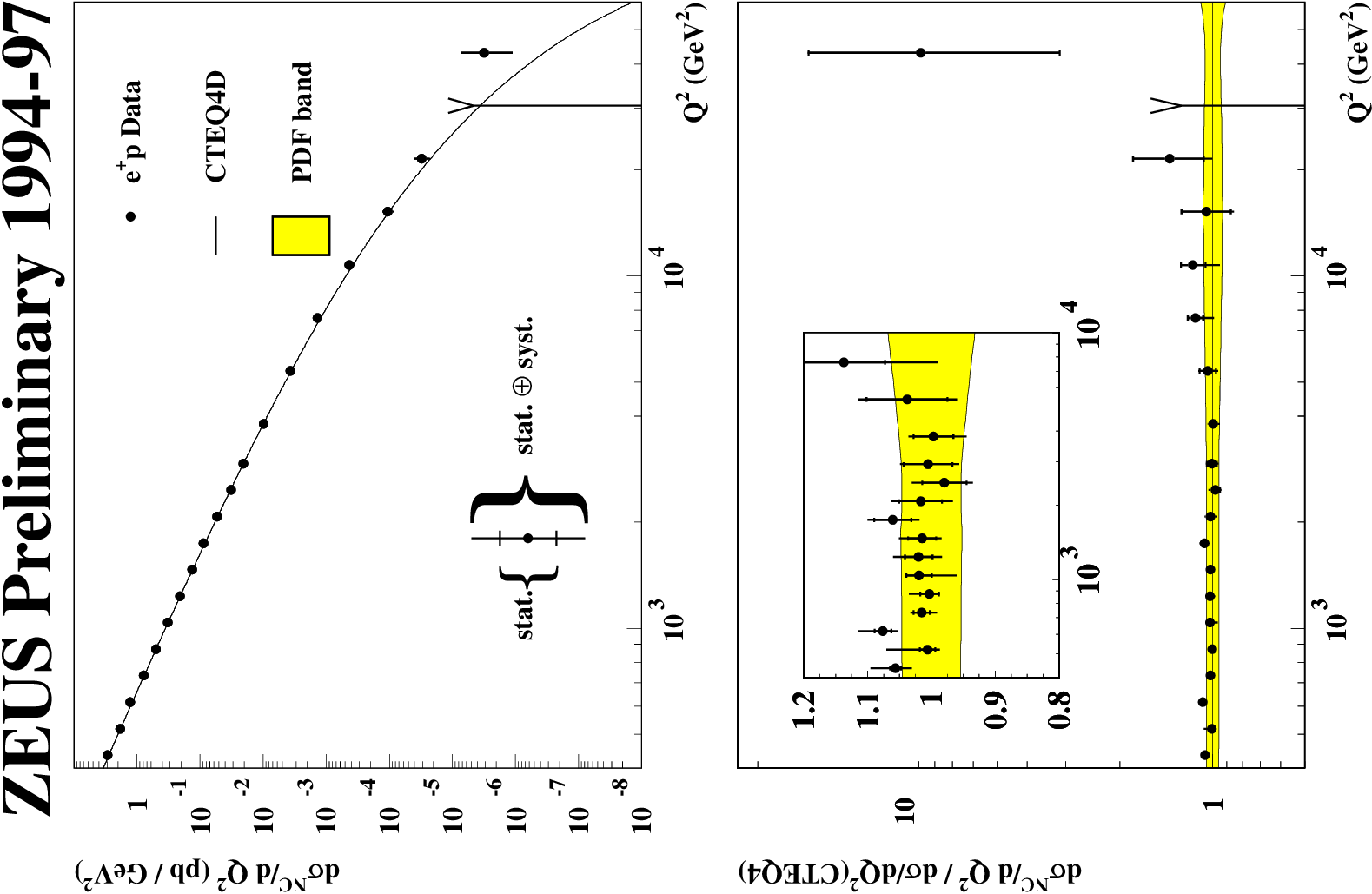}
  \hspace{12pt}
  \begin{rotate}{90}
    \parbox{0.45\textheight}{
      \caption{\label{nc_xsec}
Neutral current cross-section $d\sigma^{NC}/dQ^2$ versus $Q^2$ for $0<y<1$ 
(upper plot) and ratio 
with respect to the standard model prediction (lower plot).}}
  \end{rotate}
\end{figure}
\addtocounter{figure}{-2}%
\begin{figure}[p]
  \includegraphics[angle=90,width=0.90\textwidth]{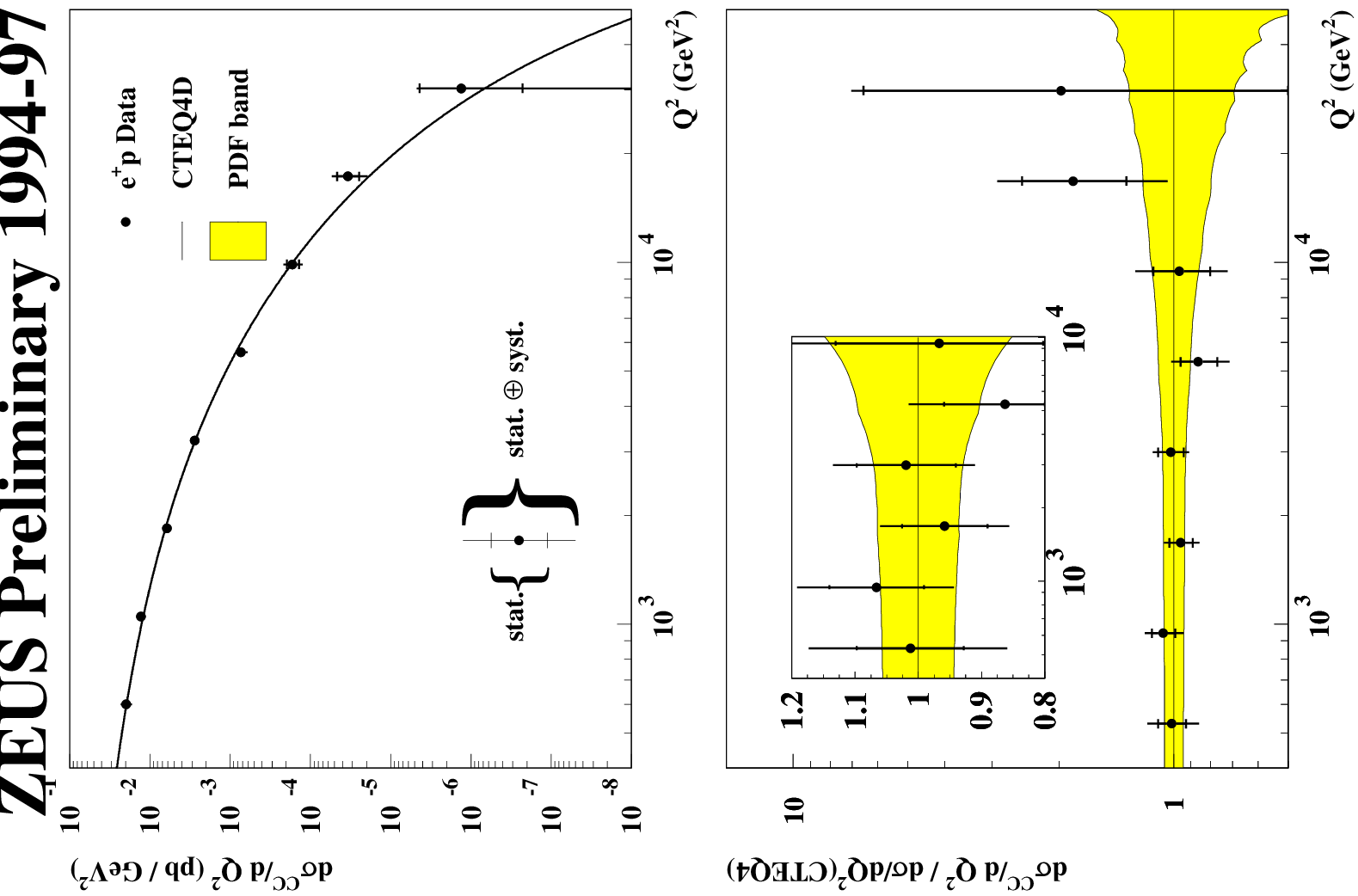}
  \hspace{12pt}
  \begin{rotate}{90}
    \parbox{0.45\textheight}{
      \caption{\label{cc_xsec}
Charged current cross-section $d\sigma^{CC}/dQ^2$ versus $Q^2$ for $0<y<1$ 
(upper plot) and ratio 
with respect to the standard model prediction (lower plot).}}
  \end{rotate}
\end{figure}

{\it Neutral Current Cross-Sections:~\cite{davew}}
High-$Q^2$ neutral current events are easily identified from the 
high-energy scattered positron.
The cross-section is particularly sensitive to the valence $u$-quark 
distribution in the proton:
$$
\frac{d^2\sigma_{e^+p}}{dx\,dQ^2} \simeq \frac{2 \pi \alpha^2}{x Q^4}
[ (1 + (1-y^2)) F_2(x, Q^2) - (1 - (1-y^2))xF_3(x, Q^2)].$$
Here, $F_2 = F_2^{em}(1+\delta_Z)$ is the generalised structure function
incorporating $\gamma + Z$ terms which is sensitive to the sum of the 
quark distributions
$(xq + x\overline{q})$ and 
$xF_3$ is the parity-violating ($Z$-contribution) term which is 
sensitive to the difference of the quark distributions
$(xq - x\overline{q})$.
$d\sigma^{NC} / dQ^2$ was measured for $Q^2 > 400$~GeV$^2$
using the double-angle method,
with an RMS ($\sigma$)
resolution on $Q^2$ of $\simeq 5\%$ ($\simeq 2-3\%$),
and cross-checked using the electron method.
Approximately 38,000 events were selected and 
the systematic uncertainties, resulting from various sources, 
are typically $\sim 2-3\%$ at low $Q^2$ increasing to $\sim 10\%$
at higher $Q^2$.
In the upper plot of Fig.~\ref{nc_xsec} the cross-section is observed
to fall over more than six orders of magnitude.
The ratio of the data to the SM, adopting the CTEQ4D PDF,
is shown in the lower plot of Fig.~\ref{nc_xsec} where
good agreement is observed up to $Q^2$ of $\simeq$~30,000~GeV$^2$. 
Comparison of the the data uncertainties 
with those from theory (shaded band) indicates that the data will 
constrain the parton densities of the proton at large-$x$.

\newpage
A wide range of new interactions would modify the NC cross-sections in a way
which can be parameterised by an effective four-fermion ($eq\rightarrow eq$)
coupling.
Given a convention for the strength of the coupling ($g^2=4\pi$), 
limits can be placed on the effective mass scale ($\Lambda$)
of these 
contact interactions.
Scalar and tensor terms are constrained by earlier
experiments and atomic parity violation experiments provide strong constraints
on various vector couplings: the relative size and sign of individual terms 
in the contact interaction amplitudes 
is therefore limited to 24 different combinations.
These contact interactions all contain 
a term proportional to $1/\Lambda^4$ which
enhances the cross-section as well as a SM interference term 
proportional to $1/\Lambda^2$ 
which can either enhance or suppress the cross-section at
intermediate $Q^2$. 
No significant deviations are found and limits on the 24 models are 
set in the range of 
$\Lambda \simeq 2-5$~TeV. These limits are competitive with, and in some
cases extend, those limits set from from hadronic cross-section measurements 
at LEP and Drell-Yan electron pair production at the Tevatron.

\section*{Conclusions}
Selected highlights from the ZEUS analyses of HERA data have been presented.
As G. Sterman noted at the DIS97 workshop~\cite{george}
``We must use the QCD we know well to investigate new physics;
but we must also pursue the QCD we do not know well."
At the DIS98
workshop a wealth of precise ZEUS data enabled us to 
identify weak links in our understanding of QCD,
to start to probe the electroweak sector and
to set limits on physics beyond the standard model.

\section*{Acknowledgements}
It is a pleasure to thank the 
organisers for an excellent workshop.
Many thanks to the members of the ZEUS collaboration listed below,
as well as to
Allen Caldwell, Peppe Iacobucci, Alex Prinias, David Saxon
and Arnulf Quadt for their help, encouragement, support and advice.


\end{document}